\documentclass[aps,prb,onecolumn,reprint,floatfix,superscriptaddress,footinbib]{revtex4}
\usepackage{xcolor}
\usepackage{microtype}
\usepackage[bookmarks=true,
bookmarksnumbered=false, 
bookmarksopen=false, 
colorlinks=true,
linkcolor=blue]{hyperref}
\definecolor{webblue}{rgb}{0, 0, 0.5} 
\hypersetup{colorlinks=true,urlcolor=blue,citecolor=blue}
\usepackage{graphicx}
\usepackage{subfigure}
\usepackage{url}
\usepackage{hyperref}
\usepackage{inconsolata}
\usepackage{amsmath}
\usepackage[capitalize]{cleveref}
\usepackage{braket}
\usepackage{outlines}
\usepackage{enumitem}
\usepackage{soul}
\usepackage{color}

\usepackage{bm} 

\usepackage[utf8]{inputenc}
\usepackage{siunitx,booktabs}
\usepackage{multirow}
\usepackage{amssymb}

\usepackage{relsize}

\usepackage{soul}
\usepackage{tabularx}

\begin{document}

\begin{abstract}
This work introduces a new class of two-dimensional crystals with the structure AC$_8$XC$_8$, consisting of two layers of graphene, a chalcogen (X = O, S, Se, Te) intercalation layer, and an alkaline earth (A = Be, Ca, Mg, Sr, Ba) adlayer. The electronic band structure for the 20 compounds was studied using density functional theory. The chalcogen $p$ orbitals interact with the carbon $\pi$ orbitals to form weakly dispersing bands that give rise to complex Fermi surfaces featuring electron and hole pockets whose densities exactly compensate each other, and van Hove singularities that are very close to, or coincident with, the Fermi level in the majority of compounds studied. The resulting electron-electron interaction effects are studied using both the temperature-flow renormalisation group approach and a spin fluctuation model, which show a dominant ferromagnetic instability coexisting with $p$- or $f$-wave spin triplet superconductivity over a range of temperatures.
\end{abstract}

\title{Superconductivity and magnetic ordering in chalcogen-intercalated graphene bilayers with charge compensation}

\author{Tommy Li}
\affiliation{Dahlem Center for Complex Quantum Systems and Fachbereich Physik, Freie Universit\"{a}t Berlin, Arnimallee 14, 14195 Berlin, Germany}
\email{tommyliphys@gmail.com}
\date{\today}
\maketitle

Superconductivity in carbon-based materials is a rapidly growing area of interest that presents both intriguing theoretical challenges and a potential for major technological advances. Having originated with the experimental discovery of superconducting states in bulk graphite intercalation compounds \cite{Hannay1965,Belash1989,Weller2005}, this field  has recently come to incorporate observations of superconductivity in a variety of low-dimensional systems \cite{Cao2018,Zondiner2020,Zhou2021,Zhou2022,Zhang2023} which are believed to be connected to anomalies in the electronic band structure, notably flat bands at magic twist angles in twisted bilayer graphene \cite{LopesdosSantos2007,BistritzerMacDonald2011}. Theoretical interest, which also encompasses possible mechanisms for superconductivity in monolayer graphene \cite{Honerkamp2008,Meng2010,Nandkishore2012,Wu2013}, is strongly motivated by the possibility of identifying physical mechanisms that may extend the superconducting phases to high temperatures such as those observed in the cuprate superconductors \cite{Bednorz1986,Subramanian1988}, and focuses on the nature of possible unconventional superconducting states, whether pairing arises due to the electron-phonon or electron-electron interaction, and the possible connections between their mechanisms of origin and those responsible for the high-$T_c$ superconductivity observed in the cuprates and iron pnictides \cite{Tseui2000,Damascelli2003,Gu2017}.

In this work I introduce a class of two-dimensional materials in which two layers of graphene are combined with lattices of chalcogen and alkaline earth atoms in separate layers. While previous experimental work on intercalation of bilayer graphene has focused on metallic intercalants, which are highly ionised and dope the carbon $\pi$ bands to high densities \cite{Ichinokura2016,Ji2019,Wang2022,Grubišić-Čabo12023,Astles2024}, in the systems I propose, the presence of a chalcogen lattice in addition to the metal ions gives rise to complex Fermi surfaces with electron and hole pockets whose densities exactly compensate each other, allowing a conducting state at net zero carrier density. The states near the Fermi level primarily move in the intercalation plane, being formed from the chalcogen $p$ orbitals, and constitute relatively flat bands hosting van Hove singularities, with a strong enhancement of the density of states near the Fermi level leading to spontaneous electronic ordering. I find that the dominant instability is ferromagnetic, and $p$- or $f$-wave spin triplet superconductivity being a subleading instability which fully coexists with ferromagnetism over a range of temperatures.

\begin{figure}
\includegraphics[width=0.9\textwidth]{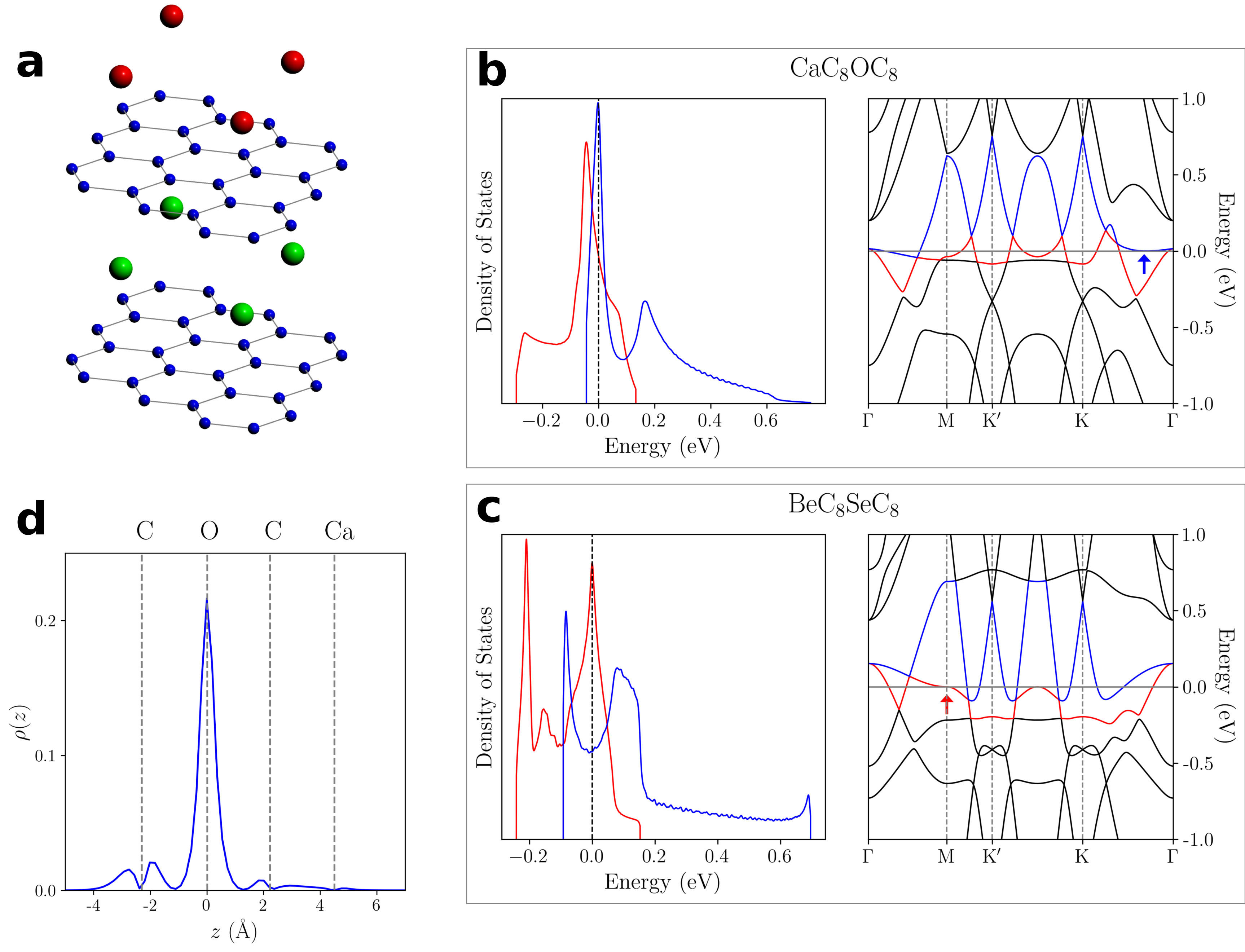}
\caption{The crystal lattice and band structure of the charge-compensated bilayer graphene compounds: (a) the real-space lattice consisting of carbon (blue), alkaline earth (red) and chalcogen (green) atoms. (b,c) the density of states of the two bands intersecting the Fermi level (left) and electronic dispersion (right) of CaC$_8$OC$_8$ and BeC$_8$SeC$_8$. The bands hosting the electron/hole pockets are plotted in blue/red, and the arrows indicate the location of the van Hove singularities.}
\end{figure}

\begin{figure}
\includegraphics[width=0.75\textwidth]{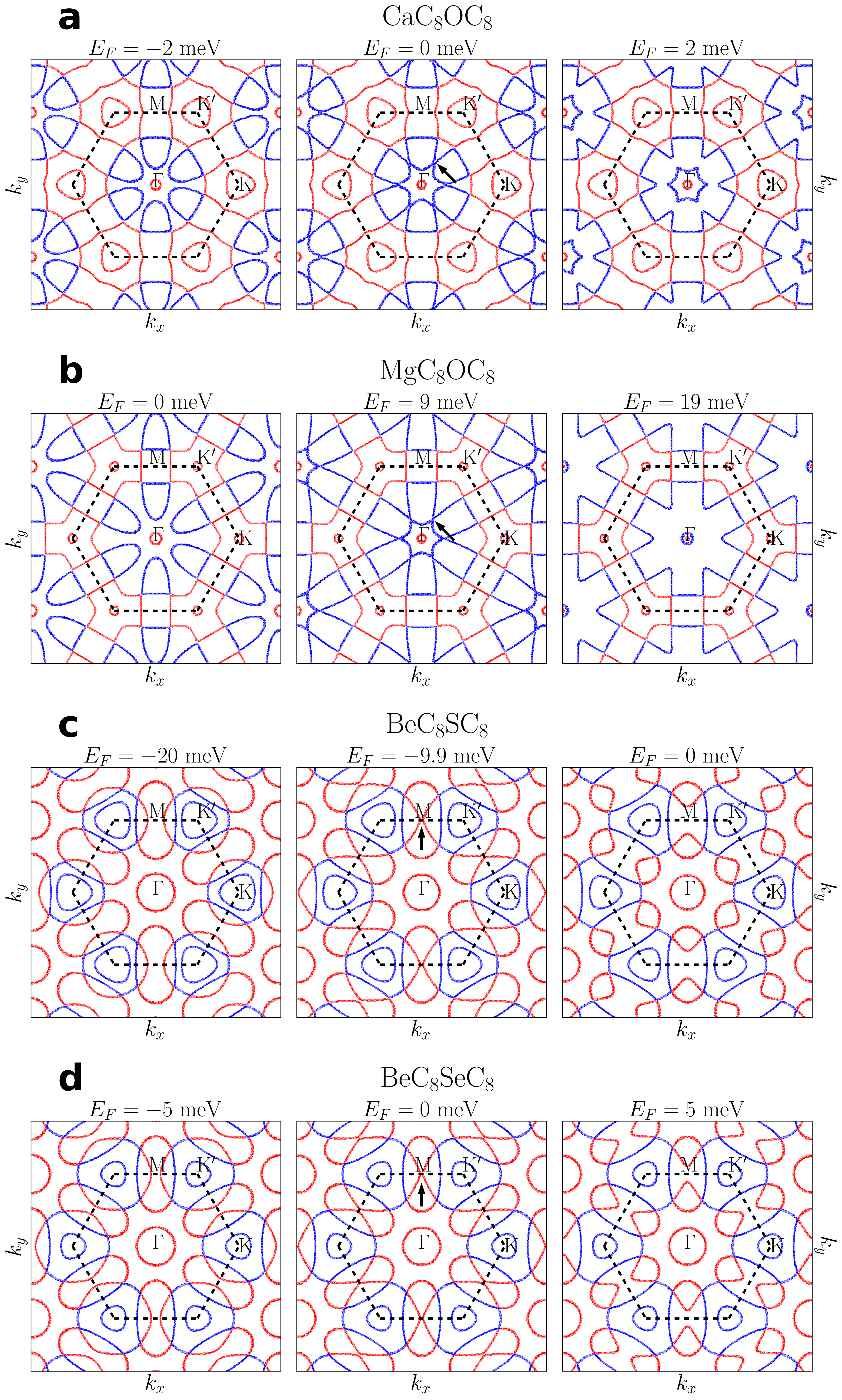}
\caption{Lifshitz transitions in (a) CaC$_8$OC$_8$, (b) MgC$_8$OC$_8$, (c) BeC$_8$OC$_8$ and (d) BeC$_8$SeC$_8$. The black arrows indicate the van Hove singularities that exist at the merging of electron or hole pockets during the transition. The boundaries of electron (hole) pockets are shown in blue (red).}
\end{figure}

The proposed crystal structure is show in Fig. 1a. Two layers of graphene (blue) are intercalated by a triangular lattice of chalcogen atoms (green) with a lattice spacing twice that of the graphene layers. A second triangular lattice, consisting of alkaline earth atoms (red) sits above the upper graphene layer. The superposition of the intercalant and adatom lattices with the honeycomb structure of graphene results in a superlattice with a unit cell containing one chalcogen atom, one alkaline earth atom, and four atoms from each of the carbon planes. The electronic band structure for the 20 compounds AC$_8$XC$_8$ with A = [Be,Mg,Ca,Sr,Ba] and B = [O,C,Se,Te] was calculated using density functional theory (DFT) with the \emph{Quantum Espresso} package \cite{Gianozzi2009} with norm-conserving, scalar relativistic pseudopotentials \cite{Hamann2013} and the DFT-D van der Waals correction \cite{Grimme2010}. An ionic relaxation calculation was performed to confirm the mechanical stability of the compounds at ambient pressure and determine the atomic positions.

The 20 compounds share strong similarities in their electronic structure. DFT results for the density of states and the electronic dispersion along a path $\Gamma$--M--K$^\prime$--K--$\Gamma$ in the folded Brillouin zone are shown in Fig. 1b and 1c for CaC$_8$OC$_8$ and BeC$_8$SeC$_8$, with results for the remaining 18 compounds provided in the Supplementary Information. The bands closest to the Fermi level are primarily shaped by the interaction between the $\pi$ bands of the carbon layers and the intercalant $p$ orbitals. Electrostatic interaction with the partially ionised chalcogen and alkaline earth layers results in an energy splitting between the $\pi$ bands of the upper and lower carbon planes, as well as a shift of the intercalant states towards the Fermi level.

As the separation between chalcogen atoms (twice the lattice constant of the carbon monolayers) is much larger than the atomic radius, electronic motion in the intercalant plane is exponentially suppressed, and the dispersion of the intercalant states results purely from hybridisation with the carbon $\pi$ bands. This hybridisation becomes strong in a small region of the Brillouin zone where small electron pockets form around the K points, but otherwise is suppressed by the large energy separation between the highly dispersing electronic states in the carbon planes and the chalcogen $p$ orbitals. As a result, the intercalant bands are an order of magnitude flatter than the $\pi$ bands in graphene.

The density of states in all 20 compounds exhibits a strong enhancement in a range of energies of the order of 100 meV surrounding the Fermi energy, reflecting the suppression of the group velocity of the intercalant states. Furthermore, logarithmic peaks appear, corresponding to saddle-point van Hove singularities which are shown in Fig 1b and 1c to coincide with the Fermi energy in CaC$_8$OC$_8$ and BeC$_8$SeC$_8$ and are indicated in the plots of the electronic dispersion by the blue and red arrows. In the oxygen- and selenium-intercalated compounds, these appear within 9 meV of the Fermi level, whereas they appear between 10 and 16 meV from the Fermi level in the sulfur-intercalated compounds and between 23 and 50 meV from the Fermi level in the tellurium-intercalated compounds. Multiple peaks in the density of states are also observed in some of the sulfur- and tellurium-intercalated compounds (see Supplementary Information). The vertical profile of the electron density for the Bloch wavefunction at the van Hove singularity $\rho(z) = \int|\psi(x,y,z)|^2 dxdy$ in CaC$_8$OC$_8$ is shown in Fig. 1d. The wavefunction is confined to the intercalant plane, with a weak admixture of carbon states, as was found for all of the compounds studied.

In all 20 compounds, the van Hove singularities are associated with Lifshitz transitions, where the topology of the Fermi surface changes as a function of the Fermi energy. These transitions are shown in Fig. 2 for CaC$_8$OC$_8$ and BeC$_8$SeC$_8$, and plots for the other 18 compounds are provided in the Supplementary Information. In the oxygen-intercalated compounds, the transitions occur at the merging of six satellite electron pockets surrounding the $\Gamma$ point into a single electron pocket. In the sulfur-, selenium- and tellurium-intercalated compounds, with the single exception of BaC$_8$TeC$_8$, they occur at the merging of pairs of hole pockets at the M point of the folded Brillouin zone. These transitions are qualitatively different to the one that occurs at doping to the M point in monolayer graphene \cite{Nandkishore2012}, where the Dirac-like pockets surrounding neighbouring K points merge. In the present case the Lifshitz transitions are driven by the chalcogen $p_x$ and $p_y$ orbitals, rather than the carbon $p_z$ orbitals as in graphene.

\begin{figure}
\includegraphics[width=0.8\textwidth]{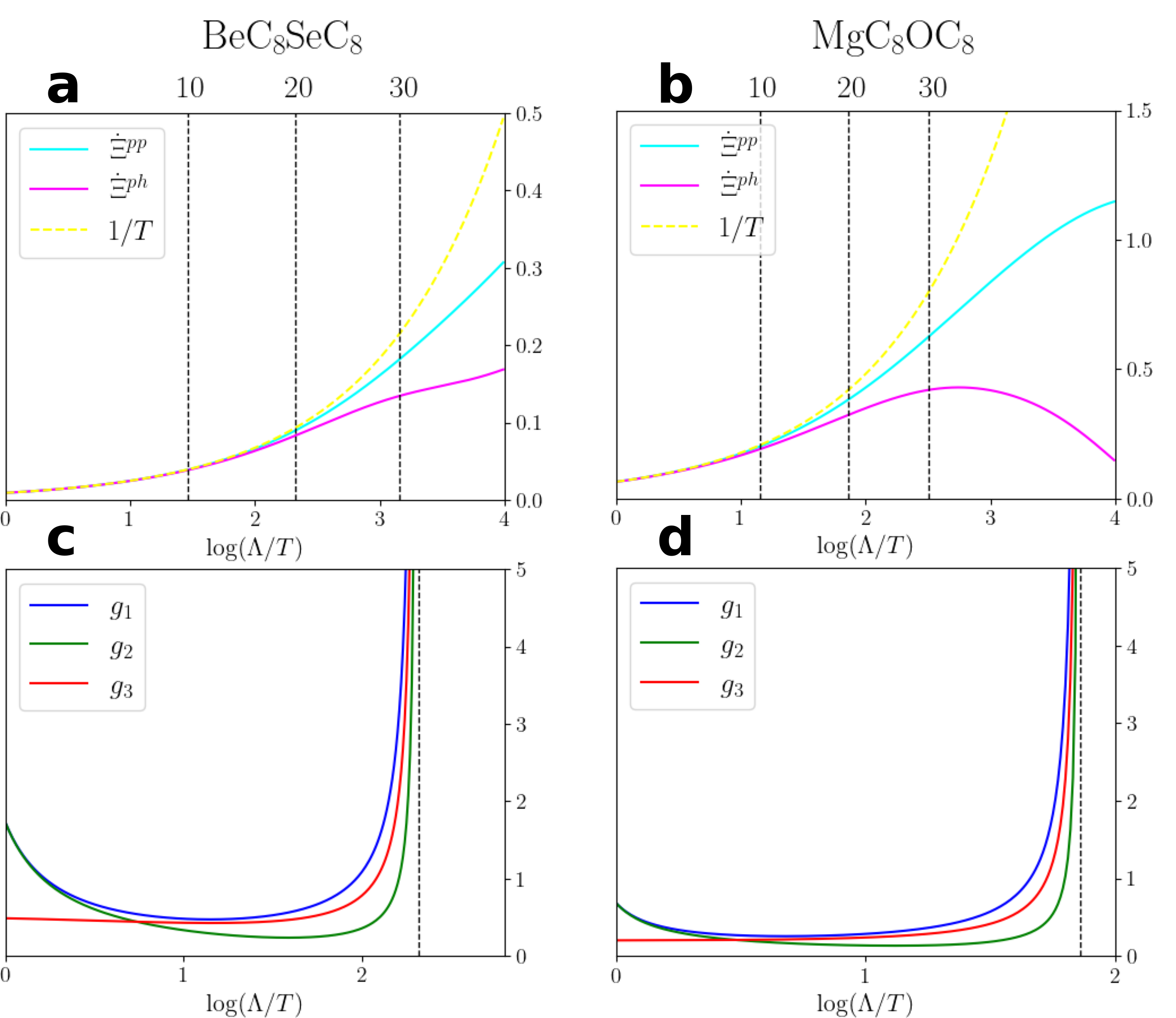}
\caption{Numerical results of the RG analysis. (a,b): the RG kernels for the particle-particle ($\dot{\Xi}^{pp}$, cyan) and particle-hole ($\dot\Xi^{ph}$, magenta) contributions to the flow equations. The approximation $\dot\Xi^{pp},\dot\Xi^{ph} \approx 1/T$ is shown in yellow, valid in the high-temperature limit. The vertical dashed lines indicate the temperatures at which the couplings diverge for $\epsilon_r = 10, 20, 30$. (c,d): The flow of the couplings $g_1,g_2,g_3$ defined in the text.}
\end{figure}

\begin{figure}
\includegraphics[width=0.7\textwidth]{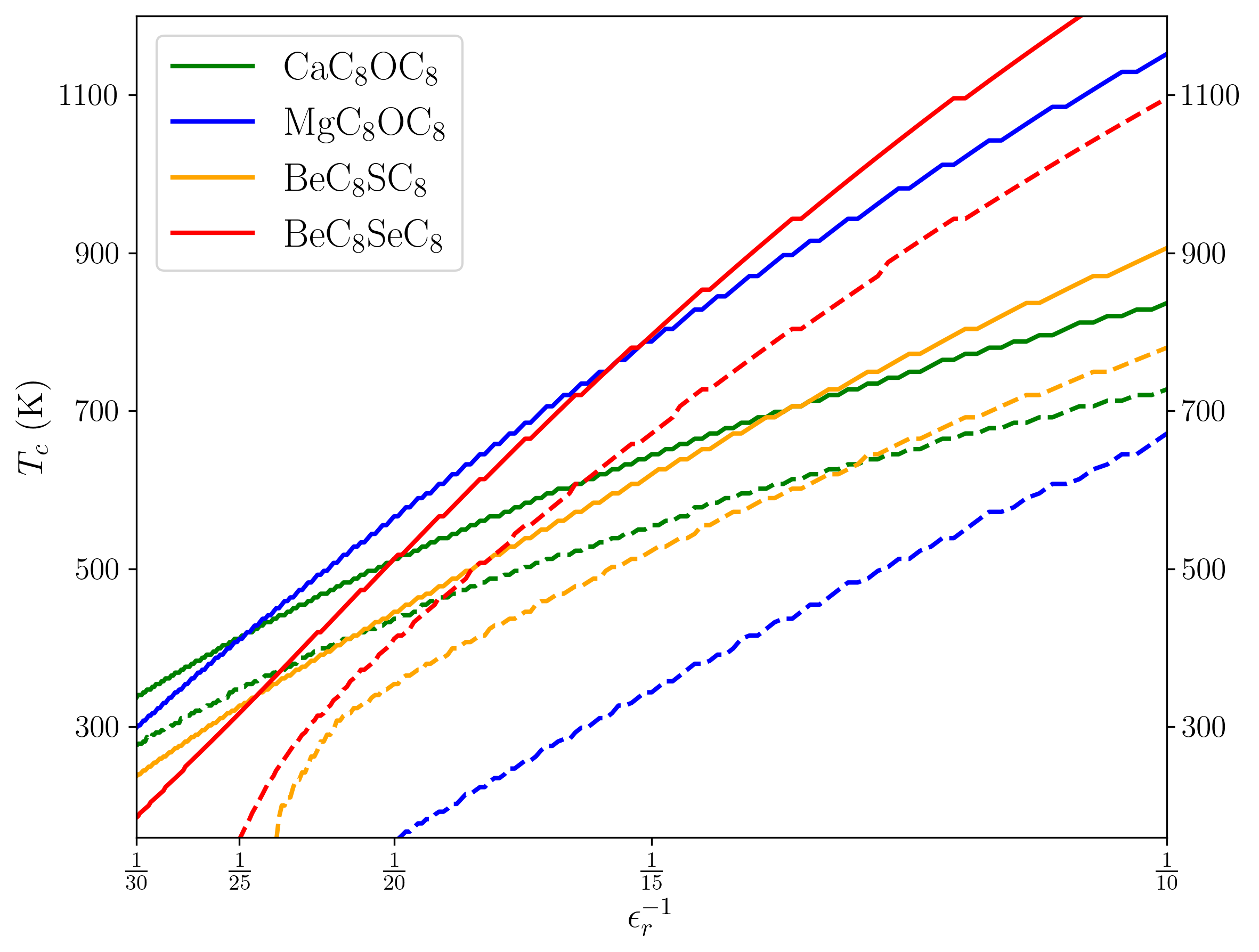}
\caption{The Curie temperatures (solid lines) and lower critical temperatures for the spatially homogeneous, equal-spin-pairing superconducting phase (dashed lines) for CaC$_8$OC$_8$, MgC$_8$OC$_8$, BeC$_8$OC$_8$ and BeC$_8$SeC$_8$. Pair density wave superconductivity with opposite spin pairing is expected to stabilise at lower temperatures.}
\end{figure}

The enhancement of the density of states as well as the presence of van Hove singularities near the Fermi level strongly suggests a possibility for correlated phases, which was explored via the temperature-flow renormalisation group, an unbiased approach that accounts for competing instabilities and allows the incorporation of the complex structure of Fermi surface \cite{Honerkamp2001}. This approach studies the temperature-dependent two-electron scattering vertex $\Gamma_T(\bm{k}_1,\bm{k}_2,\bm{k}_3,\bm{k}_4)$, which describes collisions between electrons with initial momenta $\bm{k}_1,\bm{k}_2$ and final momenta $\bm{k}_3,\bm{k}_3$ and is dressed by virtual processes that become increasingly more significant as the temperature is lowered. In order to account for the complex momentum dependence of the interactions, the Fermi surface was divided into patches, with the momentum-dependent scattering vertex replaced by a set of interaction constants $\Gamma_T(\bm{k}_1,\bm{k}_2,\bm{k}_3,\bm{k}_4) \rightarrow \Gamma_{i_1i_2i_3i_4}(T)=\Gamma_T(\bm{\kappa}_{i_1},\bm{\kappa}_{i_2},\bm{\kappa}_{i_3},\bm{\kappa}_{i_4})$ where the ``hotspots'' $\bm{\kappa}_i$ are representative momenta within each patch with a high contribution to the density of states. While the division of the Fermi surface into patches is a standard technique used in renormalisation group (RG) studies of electronic systems where multiple Fermi pockets or present, or where the momentum dependence of the collision processes plays a crucial role\cite{Maiti2010,Nandkishore2012}, the complex structure of the Fermi surfaces in the charge-compensated bilayer compounds required a much higher number of patches than previous analytical studies, ranging between 30 and 54 across the four compounds studied, CaC$_8$OC$_8$, MgC$_8$OC$_8$, BeC$_8$SC$_8$ and BeC$_8$SeC$_8$. The number of interaction constants was as high as 86058 in BeC$_8$SC$_8$, with D$_6$ symmetry allowing for these to be reduced to 6832 independent values.

The RG approach employed in this study describes the evolution of the scattering vertex with temperature due to second-order virtual processes, leading to an equation for the temperature flow of the couplings
\begin{gather}
\dot\Gamma_{i_1i_2i_3i_4} =\sum_{i_5,i_6}
\Gamma_{i_1i_2i_5i_6} \Gamma_{i_5i_6i_3i_4} \dot{\Xi}^{pp}_{i_5i_6}+ \{\Gamma_{i_1i_5i_6i_4}\Gamma_{i_2i_6i_5i_3}
-2  \Gamma_{i_1i_5i_3i_6} \Gamma_{i_2i_6i_4i_5}
 + \Gamma_{i_1i_5i_3i_6} \Gamma_{i_2i_6i_5i_4} +  \Gamma_{i_1i_5i_6i_3}  \Gamma_{i_2i_6i_4i_5}
\} \dot\Xi^{ph}_{i_5i_6}
\end{gather}
where $\dot\Gamma, \dot\Pi^{pp,ph}$ indicates differentiation with respect to the RG time $-\log T$, and $\Xi^{pp}_{i_5i_6}= \sum_{\bm{p},\bm{p}',\omega} G(\omega,\bm{p}) G(-\omega,-\bm{p}')$, $\Xi^{ph}_{i_5i_6} = -\sum_{\bm{p},\bm{p}',\omega} = G(\omega,\bm{p}) G(\omega,\bm{p}')$ with summation over momenta $\bm{p},\bm{p}'$ belong to the patches surrounding the hotspots $\bm{\kappa}_{i_5}$, $\bm{\kappa}_{i_6}$ and the fermionic Matsubara frequencies $\omega = 2\pi (n+\frac{1}{2})T$. These sums were calculated using the band structure results obtained via DFT. The initial values of the couplings were given by the matrix elements of the Coulomb interaction $V(r) = \frac{e^2}{\epsilon_r r}$ over the Bloch wavefunctions obtained from DFT, with a parameter $\epsilon_r$ accounting for screening by a dielectric substrate or due to interaction with higher bands. This was varied in the range $10\leq \epsilon_r \leq 30$ in order to study the phase diagram over a wide range of possible interaction strengths. For forward scattering processes, the matrix elements exhibit a $1/q$ divergence due to the long ranged nature of the Coulomb interaction, these were cut off by a minimum momentum transfer $q_{\text{avg}}$ representing the difference between pairs of momenta averaged over the region of the Fermi surface lying within the same patch.

The coupled flow equations were solved via numerical integration. It was found that the RG flow was dominated by contributions containing $\dot\Xi^{pp}_{ij}, \dot\Xi^{ph}_{ij}$ where $i,j$ belong to a set of oppositely positioned van Hove singularities at $\bm{\kappa}_{\text{vHS}}$ and $-\bm{\kappa}_{\text{vHS}}$. Whereas in perturbative RG, which only accounts for virtual processes that give rise to logarithmic divergences in perturbation theory, the kernels are treated as constants, in temperature-flow RG, they exhibit a non-trivial temperature dependence, which is shown in Fig. 3a and 3b for BeC$_8$SeC$_8$ and MgC$_8$OC$_8$. At high temperatures $T > T_0$ where $T_0$ is the bandwidth of the intercalant states, we find $\dot\Xi^{pp}, \dot\Xi^{ph} \sim 1/T$; as the temperature is lowered below $T_0$, the RG kernels enter a regime in which $\dot\Xi^{pp}$ scales linearly with $\log T$ while $\dot\Xi^{ph}$ saturates to a finite value. This reflects the presence of the van Hove singularity and corresponds to the temperature range in which the corrections to the scattering vertex in ordinary perturbation theory scale as $\log^2 T$ and $\log T$ in the particle-particle and particle-hole channels respectively. When the van Hove singularity is detuned from the Fermi energy by a small energy separation $\delta$, the RG kernels experience an additional transition at $T\sim \delta$ in which $\dot\Xi^{pp}$ saturates to a finite value and $\dot\Xi^{ph}$ drops to zero. This corresponds to the temperature range where the only divergent corrections to the scattering vertex in ordinary perturbation theory arise in the Cooper channel. 

The flow of the interactions $g_1 = \Gamma_T(\bm{\kappa}_{\text{vHS}},\bm{\kappa}_{\text{vHS}},\bm{\kappa}_{\text{vHS}},\bm{\kappa}_{\text{vHS}}) = \Gamma_T(-\bm{\kappa}_{\text{vHS}},-\bm{\kappa}_{\text{vHS}},-\bm{\kappa}_{\text{vHS}},-\bm{\kappa}_{\text{vHS}}), g_2 = \Gamma_T(\bm{\kappa}_{\text{vHS}},-\bm{\kappa}_{\text{vHS}},\bm{\kappa}_{\text{vHS}},-\bm{\kappa}_{\text{vHS}}), g_3 = \Gamma_T(\bm{\kappa}_{\text{vHS}},-\bm{\kappa}_{\text{vHS}},-\bm{\kappa}_{\text{vHS}},\bm{\kappa}_{\text{vHS}})$ for BeC$_8$SeC$_8$ and MgC$_8$OC$_8$ are shown in Fig. 3c and 3d. The bare values of the forward scattering amplitudes $g_1,g_2$ are much stronger than the backscattering amplitude $g_3$, which is suppressed due to the large momentum transfer. As the temperature is lowered, however, virtual processes strongly screen both $g_1$ and $g_2$, with $g_3$ remaining relatively constant in an initial phase of the RG flow, in which RPA diagrams dominate, representing the influence of the direct Coulomb interaction. At an intermediate scale, exchange contributions lead to a strong enhancement of both $g_1$ and $g_3$ which eventually dominate the flow, and all three interactions diverge at a critical temperature $T_c$, with $g_2 \ll g_1\approx g_3$. 

The divergence of the interactions also implies a divergence in the susceptibilities and therefore a destruction of the Fermi liquid phase. In order to identify the nature of the phase transitions, I considered the flow of the charge, spin and pairing susceptibilities, corresponding to perturbations of the Hamiltonian $\delta H = \sum_{\bm{k}} \Delta^{c,s}_{\bm{k},\bm{q},\mu}\psi^\dag_{\bm{k}} s^\mu \psi_{\bm{k}+\bm{q}}$ and $\delta H = \sum_{\bm{k}} \Delta^{SC}_{\bm{k},\bm{q},\mu} \psi^\dag_{\bm{k}} is^\mu s_y \psi^\dag_{\bm{q}-\bm{k}}$ with $s^\mu = \{s^0,s^x,s^y,s^z\}$. In all cases studied, the spin and superconducting susceptibilities diverge at identical critical temperatures, with the spin susceptibility showing a stronger divergence. No instability to charge ordering was observed. The leading spin instability is ferromagnetic, while the leading superconducting susceptibility is odd-momentum, spin triplet pairing, with a very weak splitting between nodal $f$-wave and chiral $p$-wave ordering, the dominant superconducting state varying between different compounds. The critical temperatures are plotted in the solid lines in Fig. 4 for the four compounds CaC$_8$OC$_8$, MgC$_8$OC$_8$, BeC$_8$SC$_8$ and BeC$_8$SeC$_8$.

Since the RG analysis only accounts for second order corrections to the vertex, with no self-energy corrections, it cannot determine whether superconductivity can coexist with ferromagnetism. To analyse this situation I consider a simplified model involving patches surrounding a single pair of van Hove singularities at $\bm{\kappa}_{\text{vHS}}$ and $-\bm{\kappa}_{\text{vHS}}$, leading to an interacting model with the three couplings $g_1,g_2,g_3$ introduced earlier. Numerically, the interaction between distinct van Hove singularities with a nonzero total momentum was observed to be weak, which justifies the two-patch model. Since the ferromagnetic instability is dominant, I perform a bosonisation of the electron-electron interaction in the exchange channel, which leads to a coupling $H_{\bm{\varphi}} = \sum_{\bm{p}\approx \bm{\kappa}_{\text{vHS}}}\psi^\dag_{\bm{p}} \left(\bm{\varphi}_+\cdot\bm{s} \right)\psi_{\bm{p}} + \sum_{\bm{p}\approx -\bm{\kappa}_{\text{vHS}}} \psi^\dag_{\bm{p}} \left(\bm{\varphi}_-\cdot\bm{s}\right) \psi_{\bm{p}}$, with the vector-valued magnetisation field $\bm{\varphi}_\pm$ having the effective potential
\begin{gather}
U\left[\bm{\varphi}_+,\bm{\varphi}_-\right] = -\frac{\alpha(T)}{2}\sum_\pm |\bm{\varphi}_\pm|^2  - b \bm{\varphi}_+\cdot\bm{\varphi}_- + \frac{\gamma(T)}{4} \sum_\pm |\bm{\varphi}_\pm|^4\ \ ,
\end{gather}
with $\alpha(T) = 2\Xi^{ph} - \frac{2g_1}{g_1^2-g_3^2}, b = \frac{2g_3}{g_1^2-g_3^2}$ and $\gamma = 2\sum_{\bm{k},\omega}(i\omega-\varepsilon_{\bm{k}})^{-4}$. The ferromagnetic phase transition occurs at the Curie temperature $T_{\text{Curie}} = T_c$ where $\alpha(T)-b$ changes sign and results in a symmetry-broken phase where $\bm{\varphi}_+,\bm{\varphi}_-$ acquire a nonzero expectation value, which we may choose to be aligned with the $z$-direction, $\langle \bm{\varphi}_+\rangle = \langle \bm{\varphi}_-\rangle = \langle \varphi\rangle \hat{\bm{z}}$.

Bosonisation of the electron-electron interaction allows us to describe the superconducting instability observed in the RG analysis as a process of pairing mediated by exchange of spin fluctuations, which is expected to occur at a temperature scale below the Curie temperature. Analysis of the superconducting gap equation shows that exchange of longitudinal spin fluctuations can mediate pairing between electrons with the same spin, leading to a condensate of Cooper pairs with a net spin polarisation parallel to the magnetic field, while transverse spin fluctuations can do so for electrons of opposite spin, leading to Cooper pairs in the $S_z = 0$ triplet state. At the Curie temperature, all spin fluctuations are gapless. Below $T_{\text{Curie}}$, the longitudinal fluctuations acquire a mass that grows as the temperature is lowered, and are eventually suppressed to the extent that the logarithmic growth of the pairing susceptibility at low temperatures is no longer sufficient to meet the criterion for pairing. This leads to a lower critical temperature for the state with equal-spin pairing, which may be estimated replacing the correlation function of spin fluctuations by the inverse mass, yielding the criterion
\begin{gather}
\langle \delta\varphi_+^z(\bm{r},t) \delta\varphi_-^z(\bm{r},t)\rangle \Xi^{pp} = \frac{b\Xi^{pp}}{4(\alpha+2b)(\alpha+b)} > 1
\label{lower-Tc}
\end{gather}
where $\delta \varphi^z_\pm = \varphi^z_\pm - \langle \varphi^z\rangle$. The lower critical temperatures for four compounds are plotted in the dashed lines in Fig. 4. For the cases of CaC$_8$OC$_8$, BeC$_8$SC$_8$, BeC$_8$SeC$_8$, the separation between the lower critical temperature and the Curie temperature is quite small, $\approx$ 100 K ($\ll T_{\text{Curie}}$), and the estimate (\ref{lower-Tc}) is not strictly justified, suggesting that the equal-spin-pairing state may be unlikely to form. For MgC$_8$OC$_8$, this separation is four times larger and the phase is more likely to exist: at temperatures close to the lower critical temperature, despite the gapping of longitudinal spin fluctuations, the resulting attractive interaction is still sufficiently strong to allow the superconducting state to form.

Transverse spin fluctuations remain gapless for all temperatures $T<T_{\text{Curie}}$. However, as the dominant instability is ferromagnetic, the exchange splitting of the Fermi surface will always be sufficiently strong to prevent pairing of opposite spins for a spatially homogeneous condensate \cite{Chandrasekhar1962,Clogston1962}. Nevertheless there exists the possibility of pairing of opposite spins in a pair density wave state \cite{Fulde1964,Larkin1965}, in which the superconducting gap exhibits periodic spatial modulation. As the gap equation depends on the correlation function for magnetic fluctuations, which contains a pole at zero frequency, a calculation of the critical temperature for this phase would require detailed knowledge of the spin wave spectrum, which I leave to future work.

Comparison of the Curie temperatures (Fig. 4) with the energy scale associated with the dispersion of the intercalant states $T_0\sim 0.1$ eV (the range of energies featuring an enhanced density of states in Fig 1b, 1c) shows that up to values of $\epsilon_r \approx 30$, the separation between the scales is within an order of magnitude. The separation grows as $\epsilon_r$ is further increased, which could be achieved by embedding the 2D system in a high-dielectric metal oxide \cite{Osada2011}. In the strong-coupling limit, the Curie temperature is expected to scale with the interaction energy $T_c\sim U$, which in the RG language corresponds to a divergence of the flow of the couplings at $T_c > T_0$. In this situation, the validity of the RG approach is severely limited. While there exists the possibility of tuning the system into the weak-coupling regime, where the approach is more justified, it would be interesting to compare the strong-coupling case to that in the cuprate and iron-based superconductors, where the majority of research into possible interactions between magnetic and superconducting instabilities is concentrated. In these systems, the superconducting phase borders on an antiferromagnetic insulator, with spin fluctuations widely considered a strong candidate for the origin of superconducting pairing \cite{Tseui2000,Damascelli2003,Gu2017}. In the systems I consider, the magnetic ordering is ferromagnetic, and while it controls the nature of the superconducting state, there is no expected insulating phase or a suppression of the density of states. The absence of a true competition between magnetic ordering and superconductivity suggests the possibility of a stronger superconducting instability than that observed in the cuprate and iron-based superconductors. Considering recent advances in metal intercalation of bilayer graphene
\cite{Ichinokura2016,Ji2019,Wang2022,Grubišić-Čabo12023,Astles2024}, the new materials proposed in this study, which likely represent an experimentally accessible extension of these developments, would provide a promising novel platform for future experimental study of the interaction between ferromagnetism and superconductivity.

I thank H.D. Scammell, K. Kolář and J. Ingham for helpful discussions.


\begin{thebibliography}{99}
\bibitem{Hannay1965} Hannay, N. B., Geballe, T. H., Matthias, B. T., Andres, K., Schmidt, P. \& MacNair, D. Superconductivity in Graphitic Compounds. \emph{Phys. Rev. Lett.} {\bf 14}, 225 (1965).  \url{https://10.1103/PhysRevLett.14.225}
\bibitem{Belash1989} Belash, I.T., Bronnikov, A.D., Zharikov, O.V. \& Pal'nichenko, A.V. Superconductivity of graphite intercalation compound with lithium C$_2$Li. \emph{Solid State Communications} {\bf 69}, (1989). \url{https://doi.org/10.1016/0038-1098(89)90933-2}
\bibitem{Weller2005} Weller, T., Ellerby, M. et al. Superconductivity in the intercalated graphite compounds C6Yb and C6Ca. \emph{Nature Phys} {\bf 1}, 39 (2005). \url{https://doi.org/10.1038/nphys0010}
\bibitem{Cao2018} Cao, Y., Fatemi, V., Fang, S. et al. Unconventional superconductivity in magic-angle graphene superlattices. \emph{Nature} {\bf 556}, 43 (2018). \url{https://doi.org/10.1038/nature26160}
\bibitem{Zondiner2020} Zondiner, U., Rozen, A., Rodan-Legrain, D. et al. Cascade of phase transitions and Dirac revivals in magic-angle graphene. \emph{Nature} {\bf 582}, 203 (2020). \url{https://doi.org/10.1038/s41586-020-2373-y}
\bibitem{Zhou2021} Zhou, H., Xie, T., Taniguchi, T. et al. Superconductivity in rhombohedral trilayer graphene. \emph{Nature} {\bf 598}, 434 (2021). \url{https://doi.org/10.1038/s41586-021-03926-0}
\bibitem{Zhou2022} Zhou, H.,  Holleis, L. et al. Isospin magnetism and spin-polarized superconductivity in Bernal bilayer graphene. \emph{Science} {\bf 375},774 (2022). \url{https://doi.org/10.1126/science.abm8386}
\bibitem{Zhang2023} Zhang, Y., Polski, R., Thomson, A. et al. Enhanced superconductivity in spin–orbit proximitized bilayer graphene. \emph{Nature} {\bf 613}, 268 (2023). \url{https://doi.org/10.1038/s41586-022-05446-x}
\bibitem{LopesdosSantos2007} Lopes dos Santos, J. M. B., Peres, N. M. R., \& Castro Neto, A. H. Graphene Bilayer with a Twist: Electronic Structure. \emph{Phys. Rev. Lett.} {\bf 99}, 256802 (2007) \url{https://doi.org/10.1103/PhysRevLett.99.256802}
\bibitem{BistritzerMacDonald2011}
Bistritzer, R. \& MacDonald, A. H. Moir\'{e} bands in twisted double-layer graphene. \emph{Proc. Natl Acad. Sci. USA} {\bf 108}, 12233 (2011).
\bibitem{Honerkamp2008} Honerkamp, C. Density Waves and Cooper Pairing on the Honeycomb Lattice. \emph{Phys. Rev. Lett.} {\bf 100}, 146404 (2008). \url{https://doi.org/10.1103/PhysRevLett.100.146404}
\bibitem{Meng2010} Meng, Z., Lang, T., Wessel, S. et al. Quantum spin liquid emerging in two-dimensional correlated Dirac fermions. \emph{Nature} {\bf 464}, 847 (2010). \url{https://doi.org/10.1038/nature08942}
\bibitem{Nandkishore2012} Nandkishore, R., Levitov, L. \& Chubukov, A. Chiral superconductivity from repulsive interactions in doped graphene. \emph{Nature Phys} {\bf 8}, 158 (2012). \url{https://doi.org/10.1038/nphys2208}
\bibitem{Wu2013} Wu, W., Scherer, M. M., Honerkamp, C. \& Le Hur, K. Correlated Dirac particles and superconductivity on the honeycomb lattice. \emph{Phys. Rev. B} {\bf 87}, 094521 (2013). \url{https://doi.org/10.1103/PhysRevB.87.094521}
\bibitem{Bednorz1986} Bednorz, J.G. \& M\"{u}ller, K.A. Possible high $T_c$ superconductivity in the Ba-La-Cu-O system. \emph{Z. Physik B - Condensed Matter} {\bf 64}, 189 (1986). \url{https://doi.org/10.1007/BF01303701}
\bibitem{Subramanian1988} Subramanian, M. A., Torardi, C.C. et al. A New High-Temperature Superconductor: Bi$_2$Sr$_{3-x}$Ca$_x$ Cu$_{2O8+y}$. \emph{Science} {\bf 239}, 1015 (1988). \url{https://doi.org/10.1126/science.239.4843.1015}
\bibitem{Tseui2000} Tsuei, C. C., Kirtley \& J. R. Pairing symmetry in cuprate superconductors. \emph{Rev. Mod. Phys.} {\bf 72}, 969 (2000). \url{https://doi.org/10.1103/RevModPhys.72.969}
\bibitem{Damascelli2003} Damascelli, A., Hussain, Z. \& Shen, Z.-X. Angle-resolved photoemission studies of the cuprate superconductors. \emph{Rev. Mod. Phys.} {\bf 75}, 473 (2003). \url{https://doi.org/10.1103/RevModPhys.75.473}
\bibitem{Gu2017} Gu, Y., Liu, Z. et al. Unified Phase Diagram for Iron-Based Superconductors. \emph{Phys. Rev. Lett.} {\bf 119}, 157001 (2017). \url{https://doi.org/10.1103/PhysRevLett.119.157001}
\bibitem{Ichinokura2016} Ichinokura, S., Sugawara, K. et. al. Superconducting Calcium-Intercalated Bilayer Graphene. \emph{ACS Nano} {\bf 10}, 2761 (2016). \url{https://doi.org/10.1021/acsnano.5b07848}
\bibitem{Ji2019} Ji, K., Han, J., Hirata, A. et al. Lithium intercalation into bilayer graphene. \emph{Nat Commun} {\bf 10}, 275 (2019). \url{https://doi.org/10.1038/s41467-018-07942-z}
\bibitem{Wang2022} Wang, X., Liu, N. et. al. Strong Coupling Superconductivity in Ca-Intercalated Bilayer Graphene on SiC. \emph{Nano Lett.} {\bf 22}, 7651 (2022). \url{https://doi.org/10.1021/acs.nanolett.2c02804}
\bibitem{Grubišić-Čabo12023} Grubišić-Čabo1, A., Kotsakidis, J. C. et. al. Quasi-freestanding AA-stacked bilayer graphene induced by calcium intercalation of the graphene-silicon carbide interface, \emph{Front. Nanotechnol.} {\bf 5} (2023). \url{https://doi.org/10.3389/fnano.2023.1333127}
\bibitem{Astles2024} Astles, T., McHugh, J.G., Zhang, R. et al. In-plane staging in lithium-ion intercalation of bilayer graphene. \emph{Nat Commun} {\bf 15}, 6933 (2024). \url{https://doi.org/10.1038/s41467-024-51196-x}
\bibitem{Gianozzi2009} Giannozzi, P. et al. QUANTUM ESPRESSO: a modular and open-source software project for quantum simulations of materials. \emph{J. Phys.: Condens. Matter} {\bf 21}, 395502 (2009). \url{https://doi.org/10.1088/0953-8984/21/39/395502}
\bibitem{Hamann2013} Hamann, D. R. Optimized norm-conserving Vanderbilt pseudopotentials. \emph{Phys. Rev. B} {\bf 88}, 085117 (2013). \url{https://doi.org/10.1103/PhysRevB.88.085117}
\bibitem{Grimme2010} Grimme, S., Antony, J., Ehrlich, S. \& Krieg, H. A consistent and accurate ab initio parametrization of density functional dispersion correction (DFT-D) for the 94 elements H--Pu. \emph{J. Chem. Phys.} {\bf 132}, 154104 (2010). \url{https://doi.org/10.1063/1.3382344}
\bibitem{Honerkamp2001} Honerkamp, C. \& Salmhofer, M. Temperature-flow renormalization group and the competition between superconductivity and ferromagnetism, \emph{Phys. Rev. B} {\bf 64}, 184516 (2001). \url{https://doi.org/doi/10.1103/PhysRevB.64.184516}
\bibitem{Maiti2010} Maiti, S. \& Chubukov, A. V., Renormalization group flow, competing phases, and the structure of superconducting gap in multiband models of iron-based superconductors. \emph{Phys. Rev. B} {\bf 82}, 214515 (2010). \url{https://doi.org/10.1103/PhysRevB.82.214515}
\bibitem{Chandrasekhar1962} Chandrasekhar, B. S. A note on the maximum critical field of high-field superconductors. \emph{Appl. Phys. Lett.} {\bf 1}, 7 (1962). \url{https://doi.org/10.1063/1.1777362}
\bibitem{Clogston1962} Clogston, A. M. Upper Limit for the Critical Field in Hard Superconductors. \emph{Phys. Rev. Lett.} {\bf 9}, 266 (1962). \url{https://doi.org/10.1103/PhysRevLett.9.266}
\bibitem{Fulde1964}
Fulde, P. \& Ferrell, R. A. Superconductivity in a Strong Spin-Exchange Field, \emph{Phys. Rev.} {\bf 135}, A550 (1964). \url{https://doi.org/10.1103/PhysRev.135.A550}
\bibitem{Larkin1965} Larkin, A. I. \& Ovchinnikov, Y. N. Nonuniform state of superconductors. \emph{Sov. Phys. JETP} {\bf 20}, 762 (1965).
\bibitem{Osada2011} Osada, M., \& Sasaki, T. Two-Dimensional Dielectric Nanosheets: Novel Nanoelectronics From Nanocrystal Building Blocks. \emph{Advanced Materials}, {\bf 24}, 210 (2011). \url{https://doi.org/10.1002/adma.201103241}
\end{thebibliography}
\end{document}


\title{Superconductivity and magnetic ordering in chalcogen-intercalated graphene bilayers with charge compensation: Supplementary Information}
\maketitle

\section*{DFT results for the compounds AC$_8$XC$_8$}

This section presents DFT results for the vertical profile of the Bloch wavefunctions at the van Hove singularity $\rho(z) = \int |\psi(x,y,z)|^2 dx dy$, the electronic dispersion, density of states and Fermi surfaces for the compounds AC$_8$XC$_8$ with A=[Be,Mg,Ca,Sr,Ba] and X=[O,S,Se,Te] not presented in the main text.

\begin{figure}[h]
\includegraphics[width=0.9\textwidth]{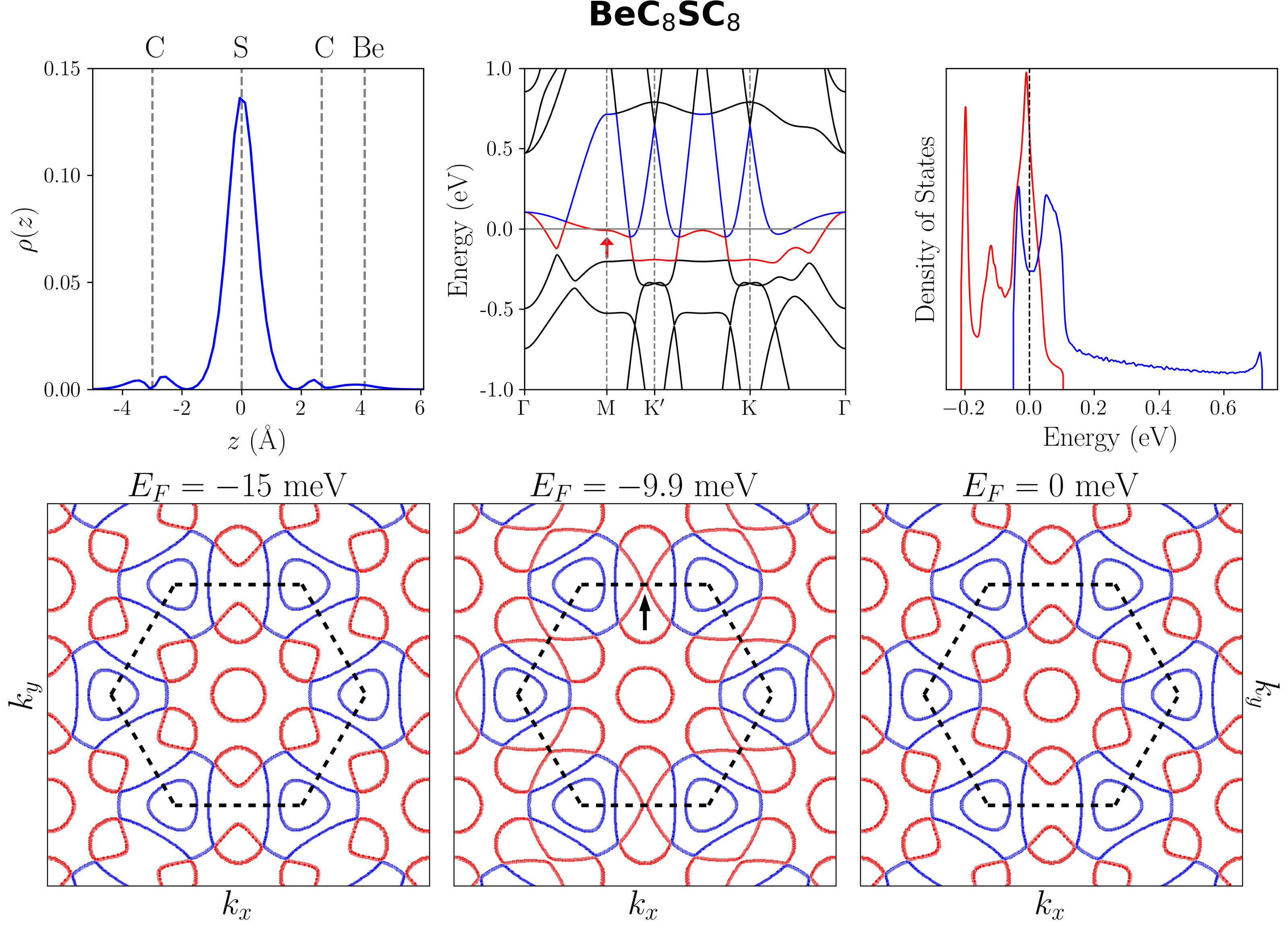}
\end{figure}

\begin{figure}[h]
\begin{tabular}{c}
\includegraphics[width=0.9\textwidth]{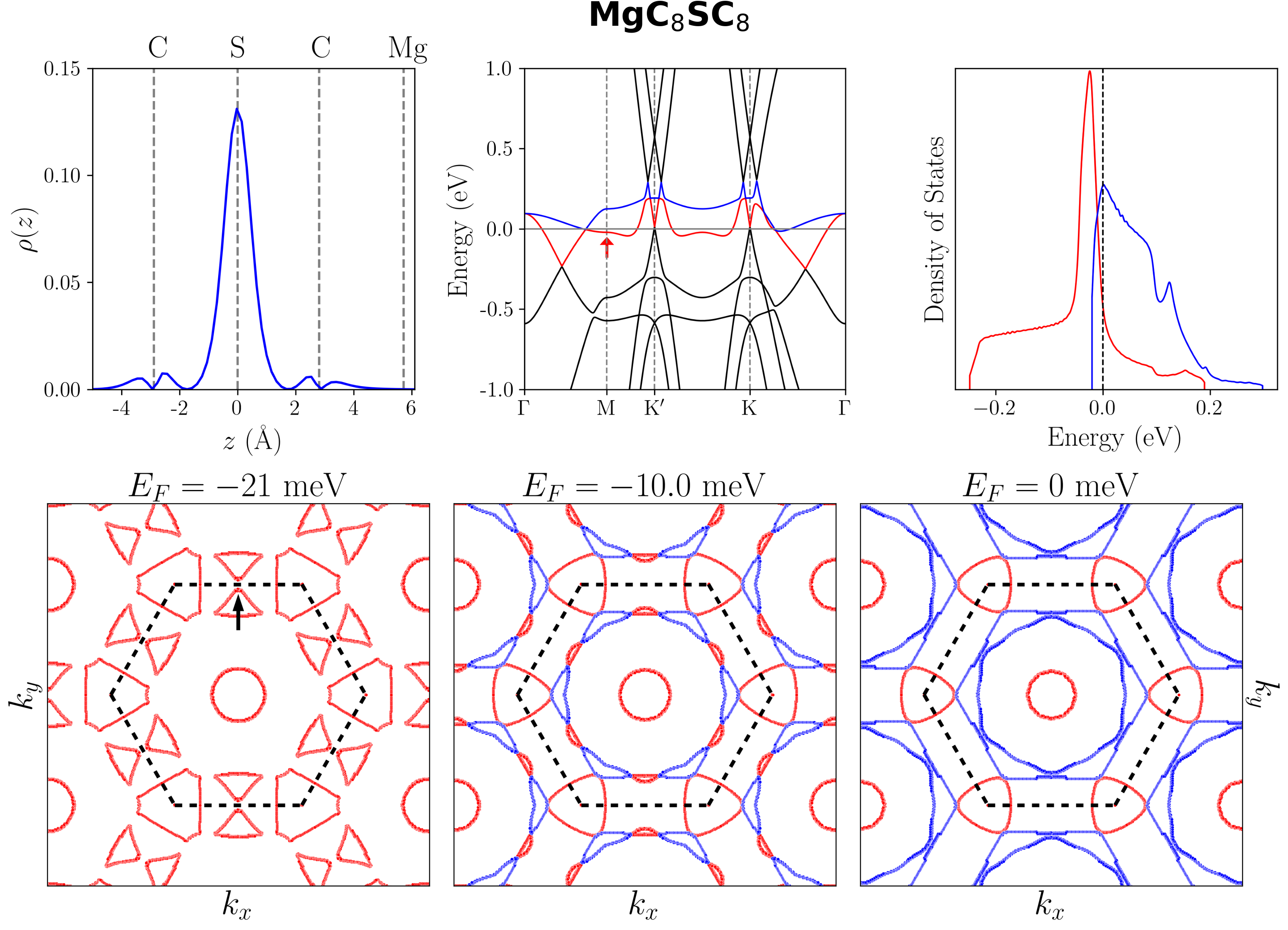}
\\
\includegraphics[width=0.9\textwidth]{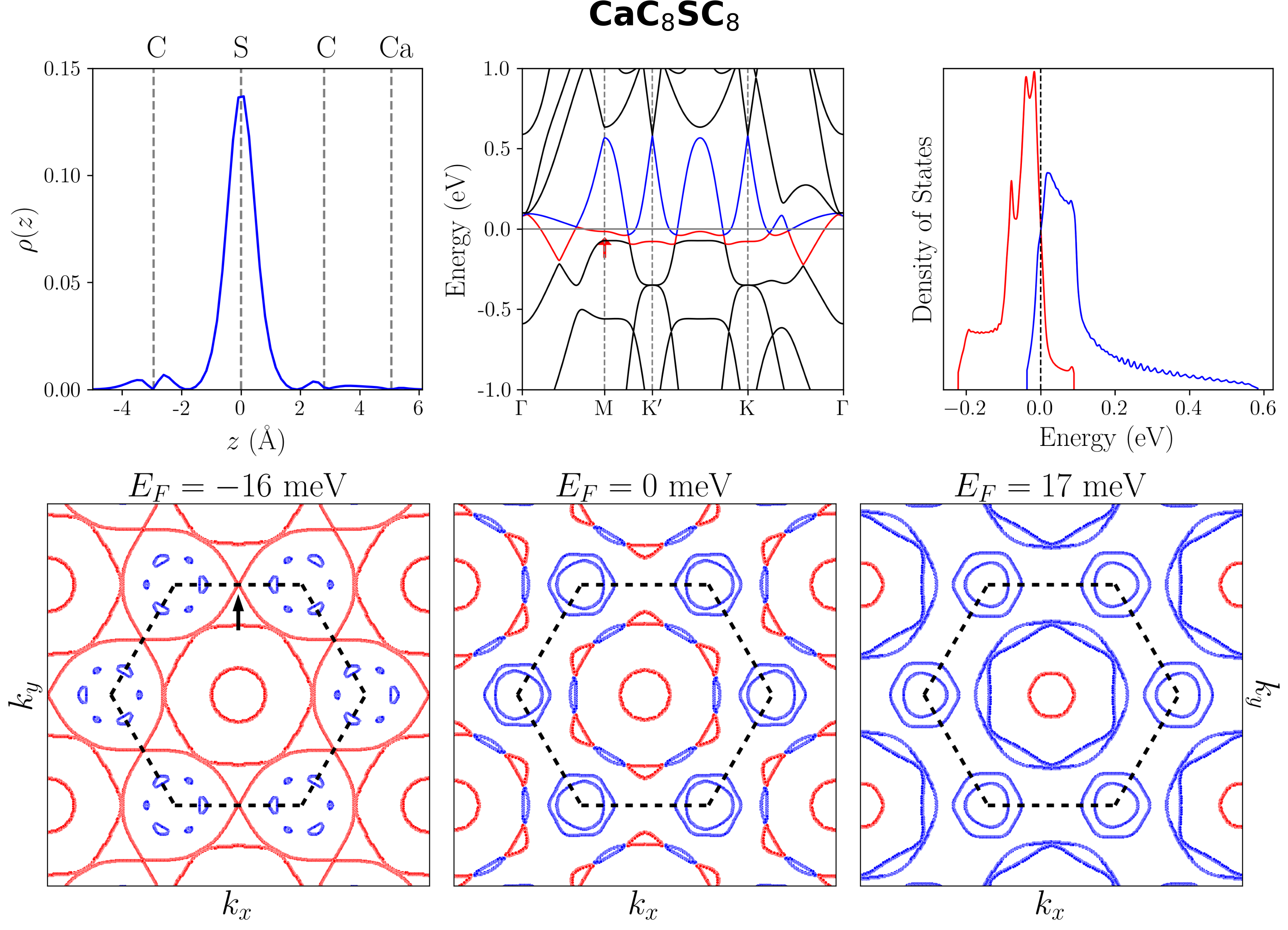}
\end{tabular}
\end{figure}

\begin{figure}[h]
\begin{tabular}{c}
\includegraphics[width=0.9\textwidth]{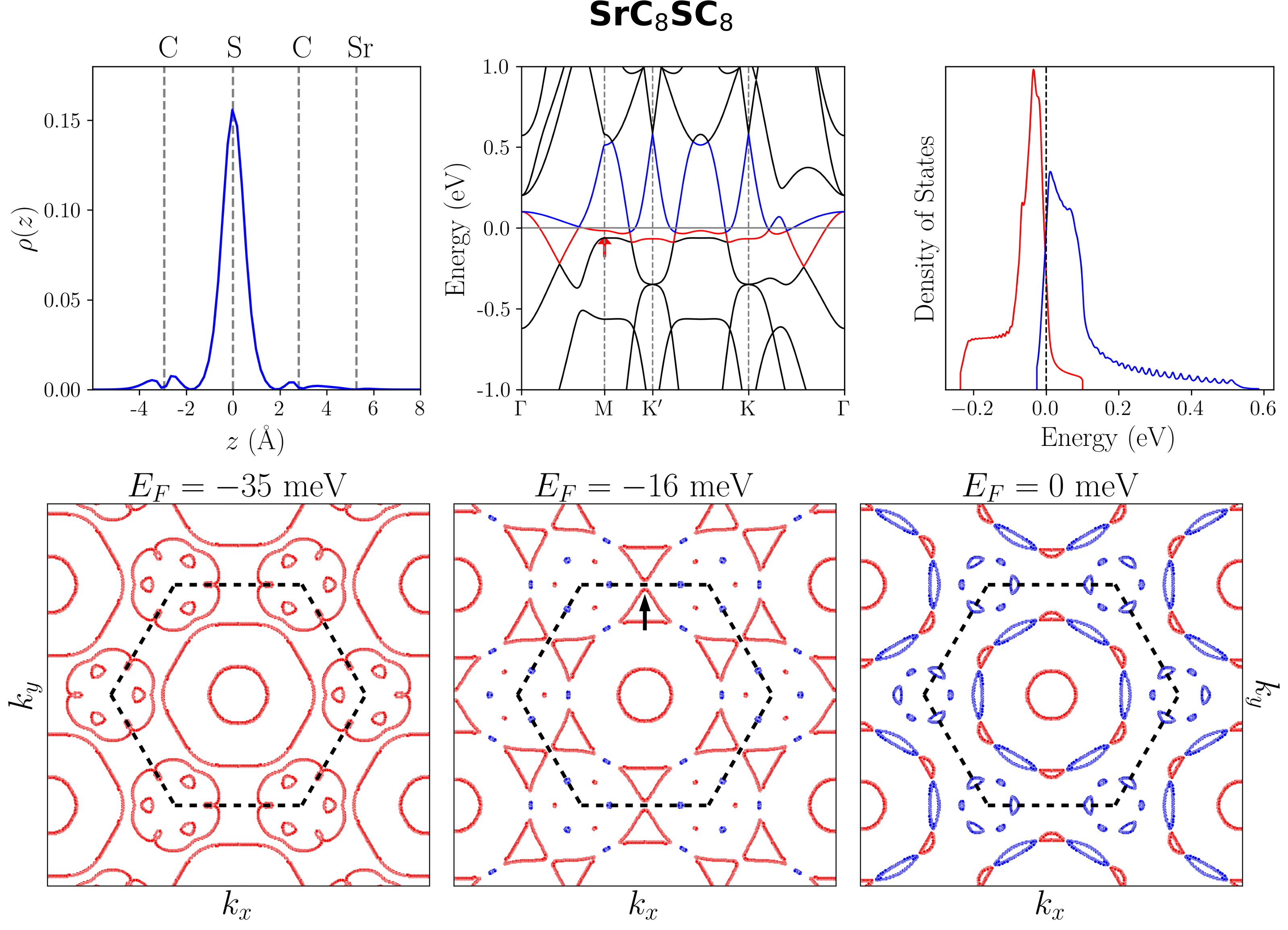}
\\
\includegraphics[width=0.9\textwidth]{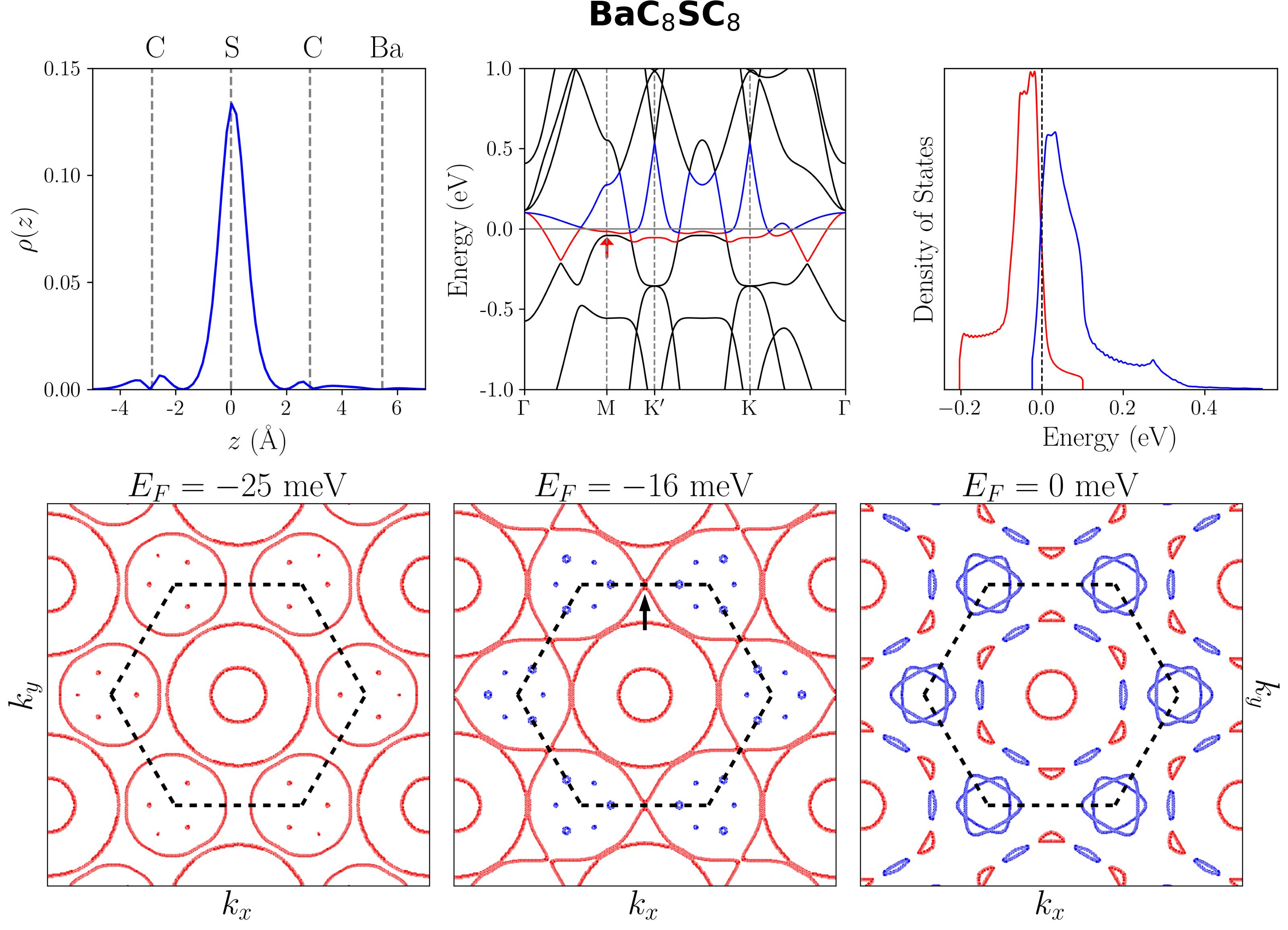}
\end{tabular}
\end{figure}

\begin{figure}[h]
\begin{tabular}{c}
\includegraphics[width=0.9\textwidth]{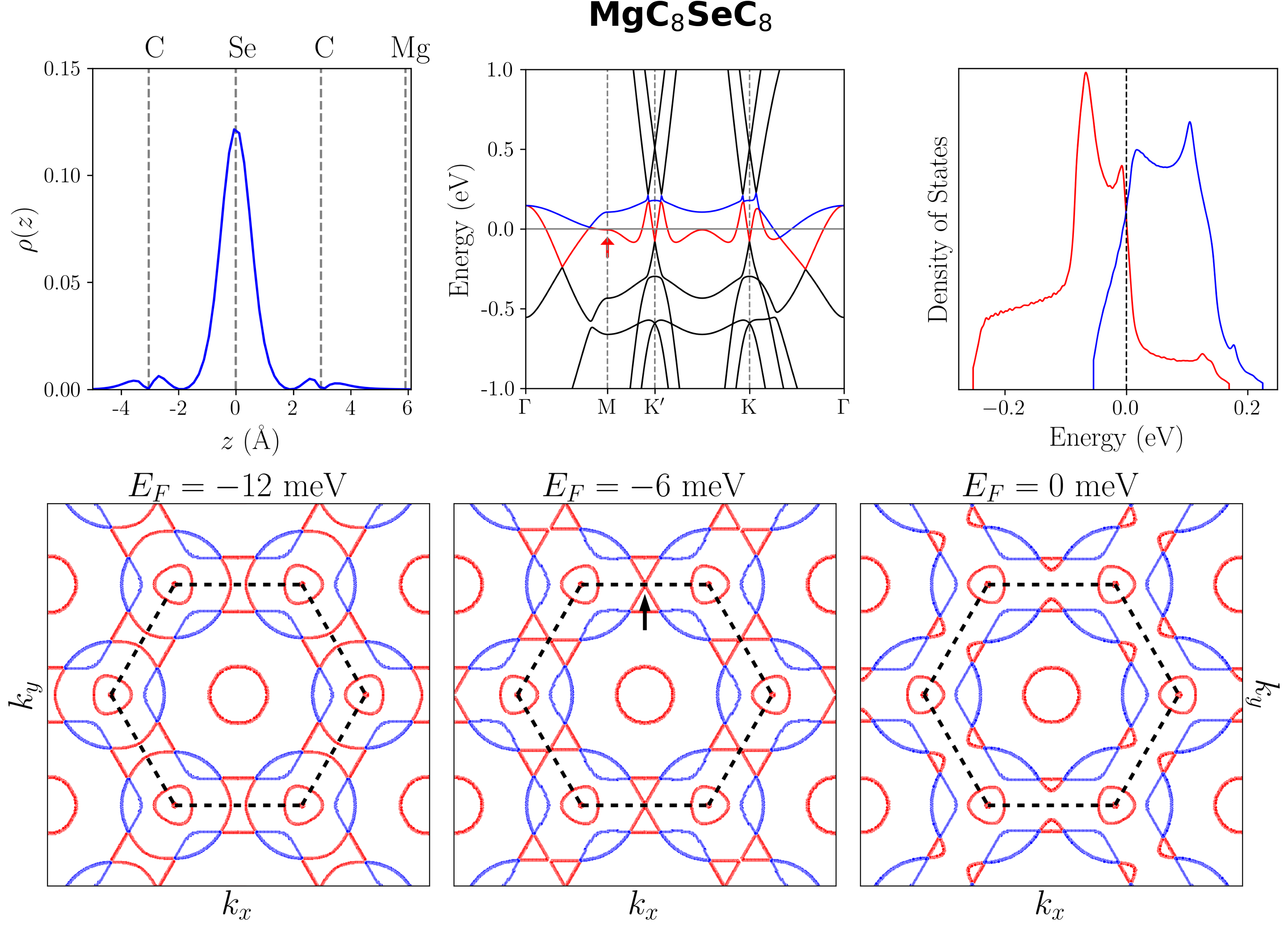}
\\
\includegraphics[width=0.9\textwidth]{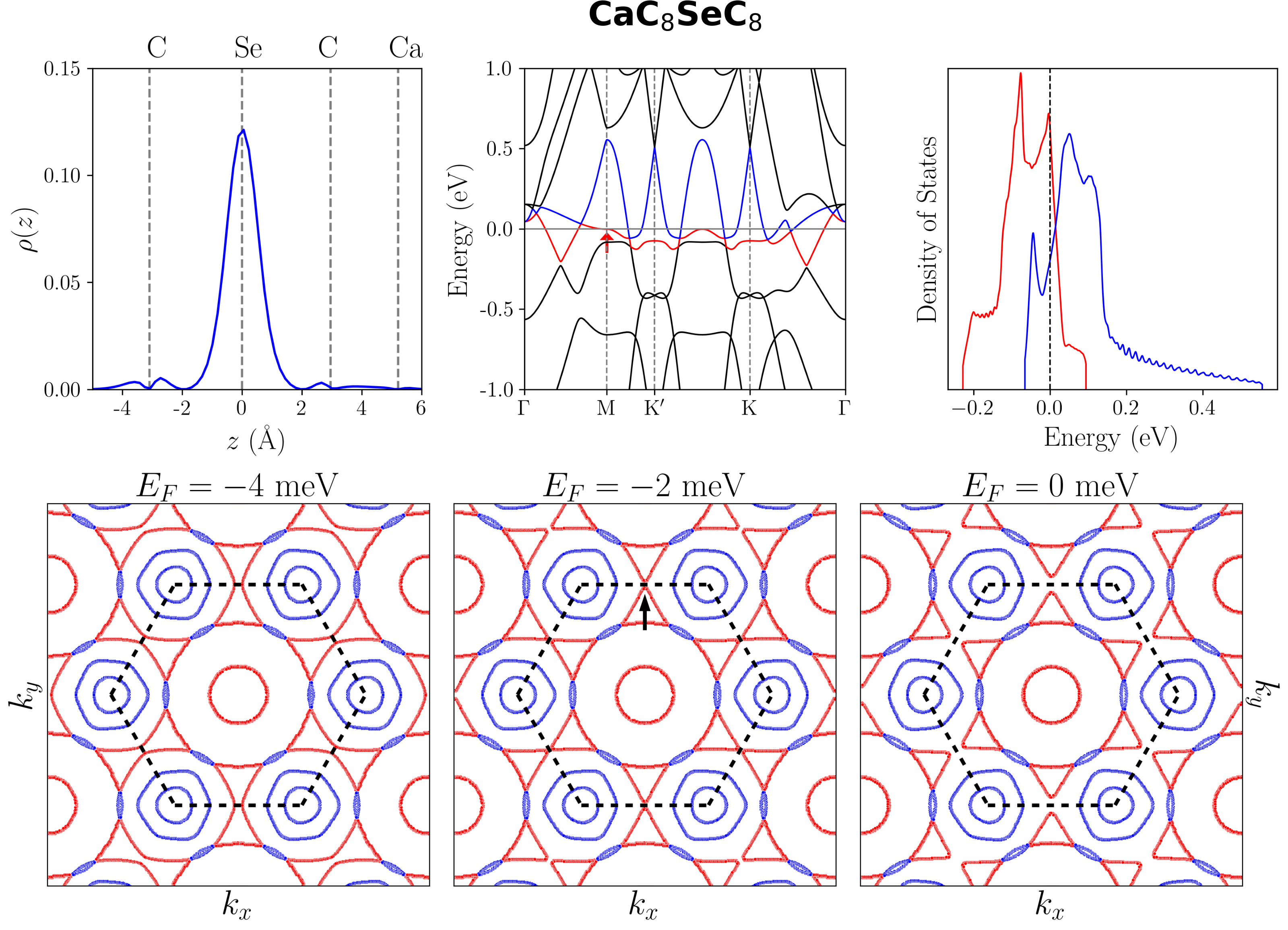}
\end{tabular}
\end{figure}

\begin{figure}[h]
\begin{tabular}{c}
\includegraphics[width=0.9\textwidth]{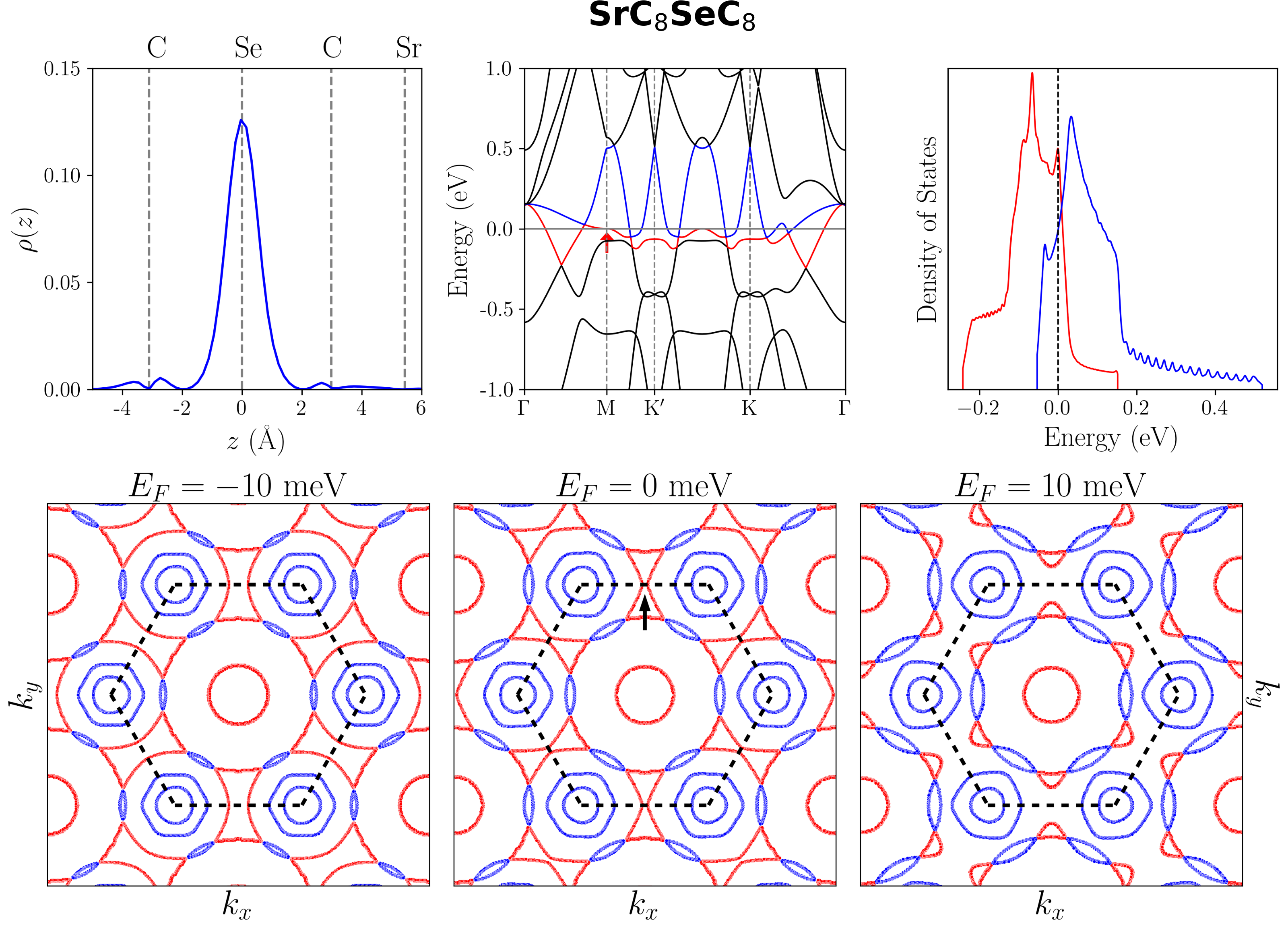}
\\
\includegraphics[width=0.9\textwidth]{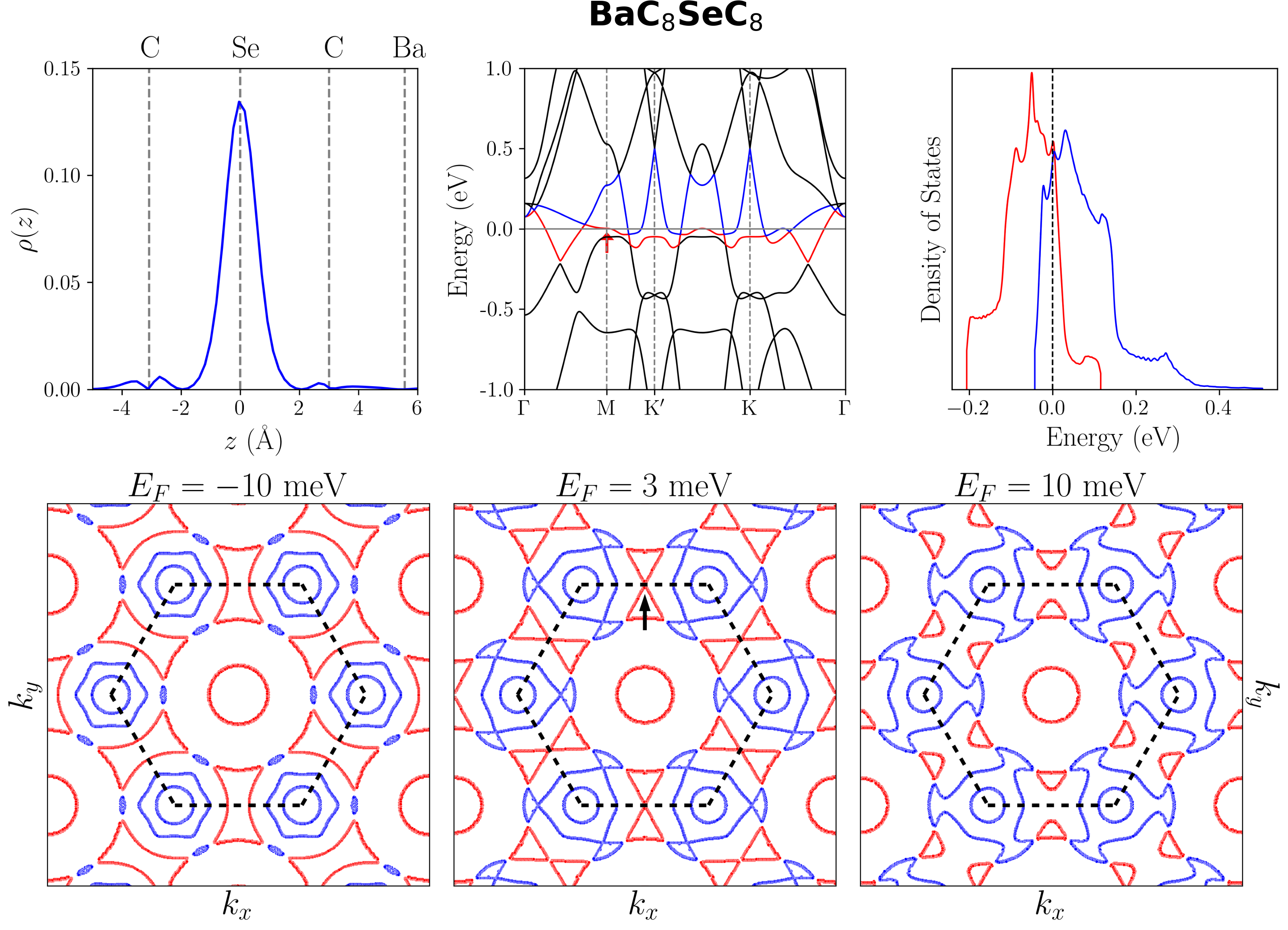}
\end{tabular}
\end{figure}

\begin{figure}[h]
\begin{tabular}{c}
\includegraphics[width=0.9\textwidth]{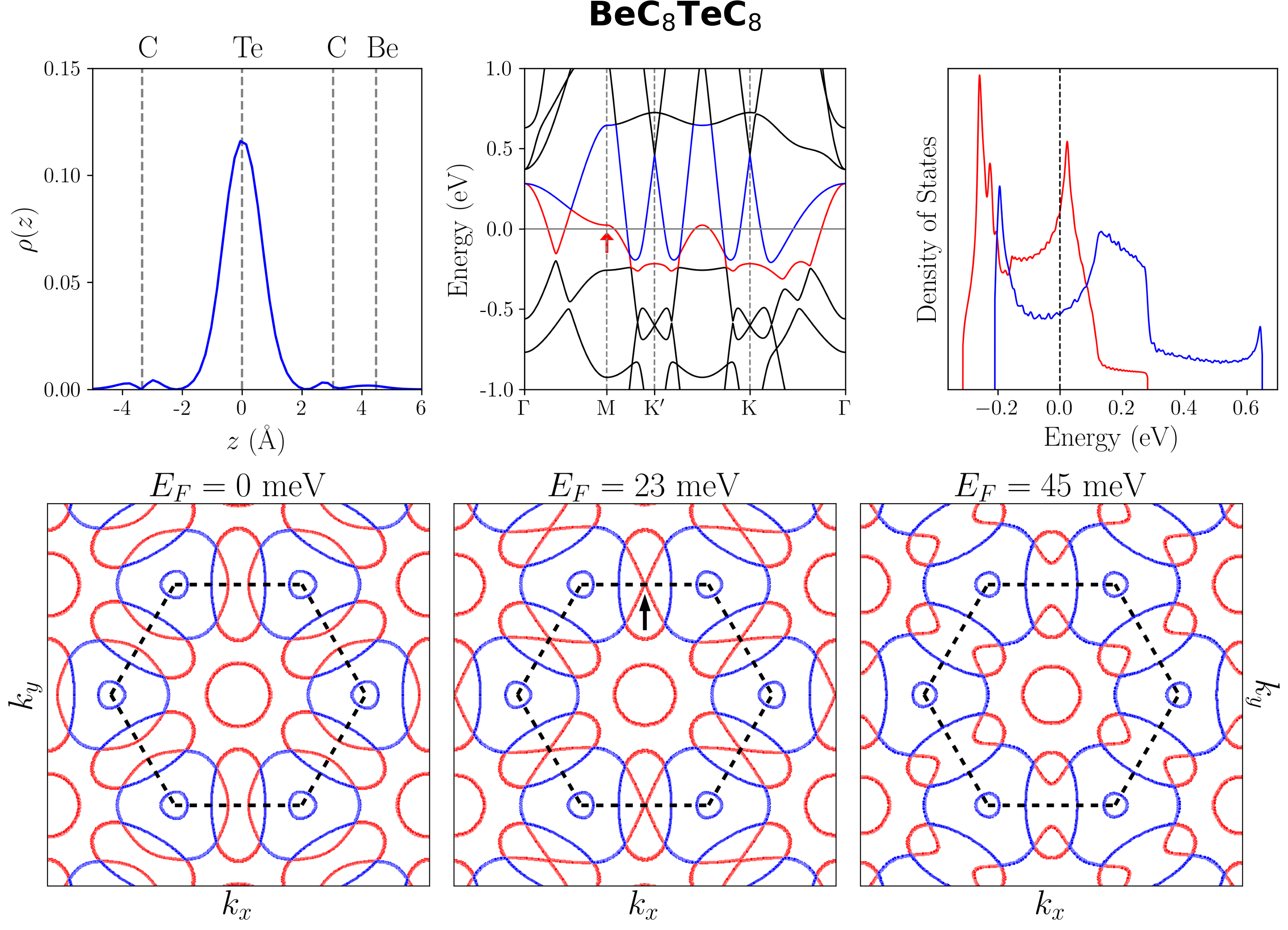}
\\
\includegraphics[width=0.9\textwidth]{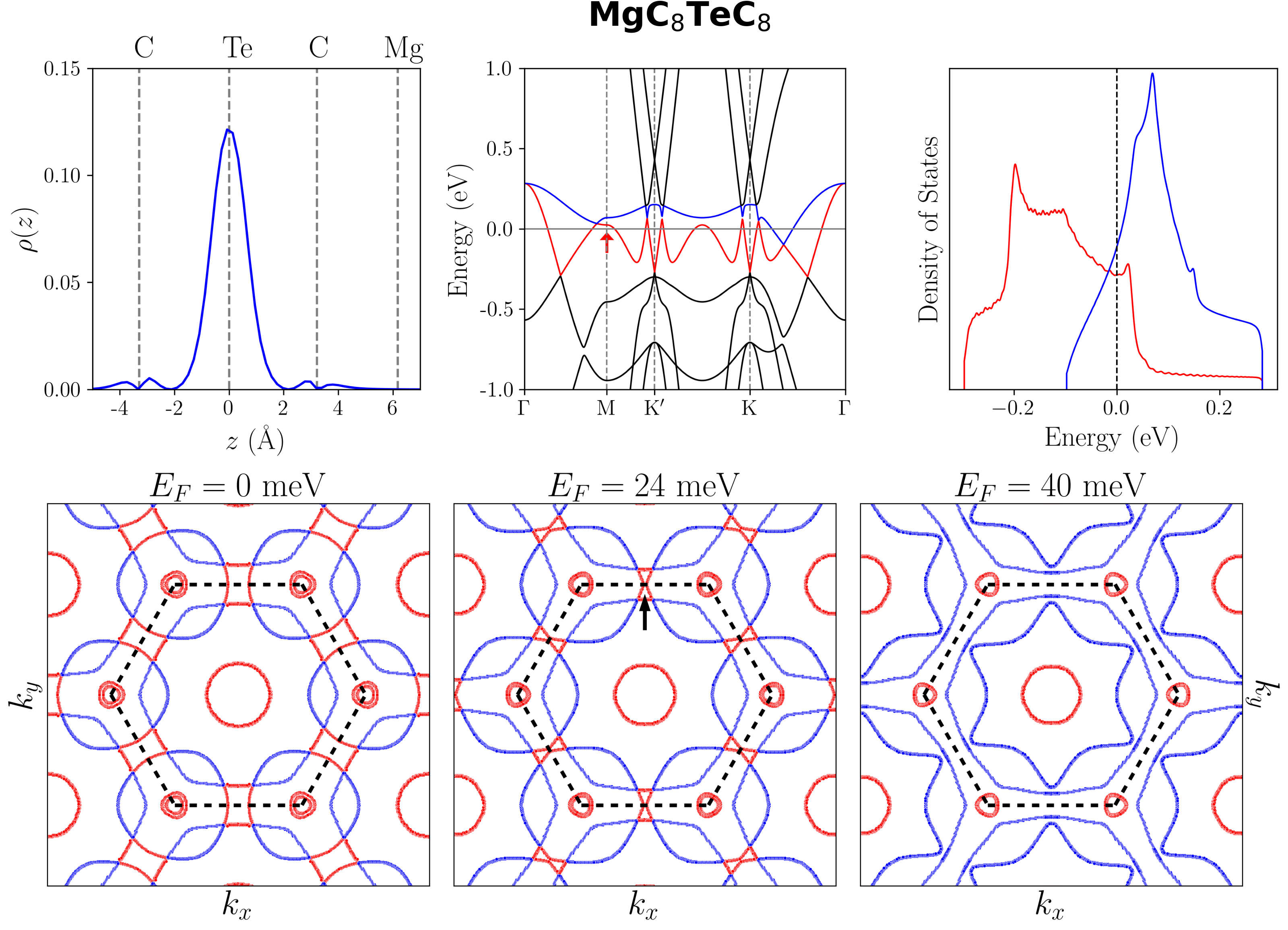}
\end{tabular}
\end{figure}

\begin{figure}[h]
\begin{tabular}{c}
\includegraphics[width=0.9\textwidth]{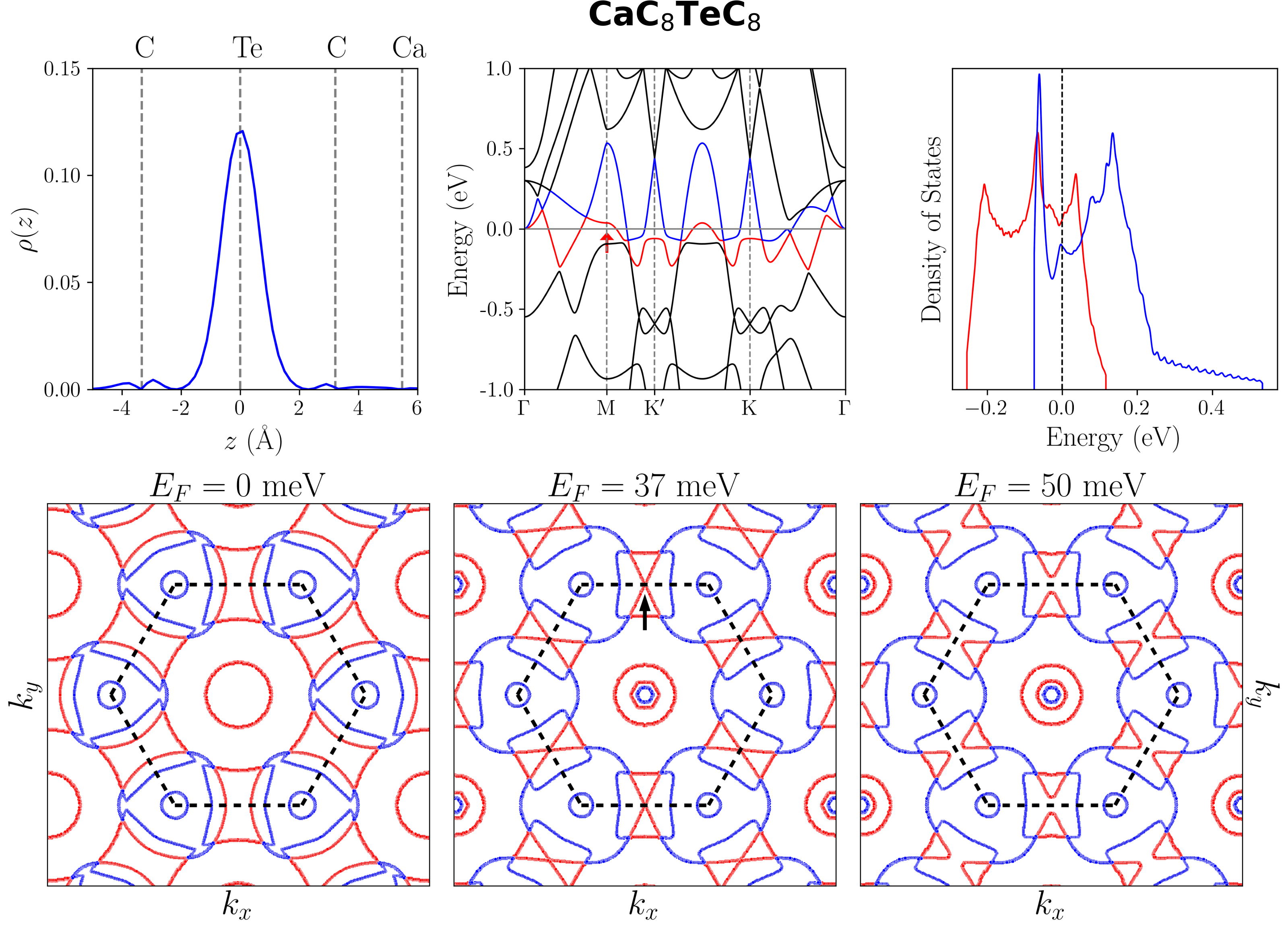}
\\
\includegraphics[width=0.9\textwidth]{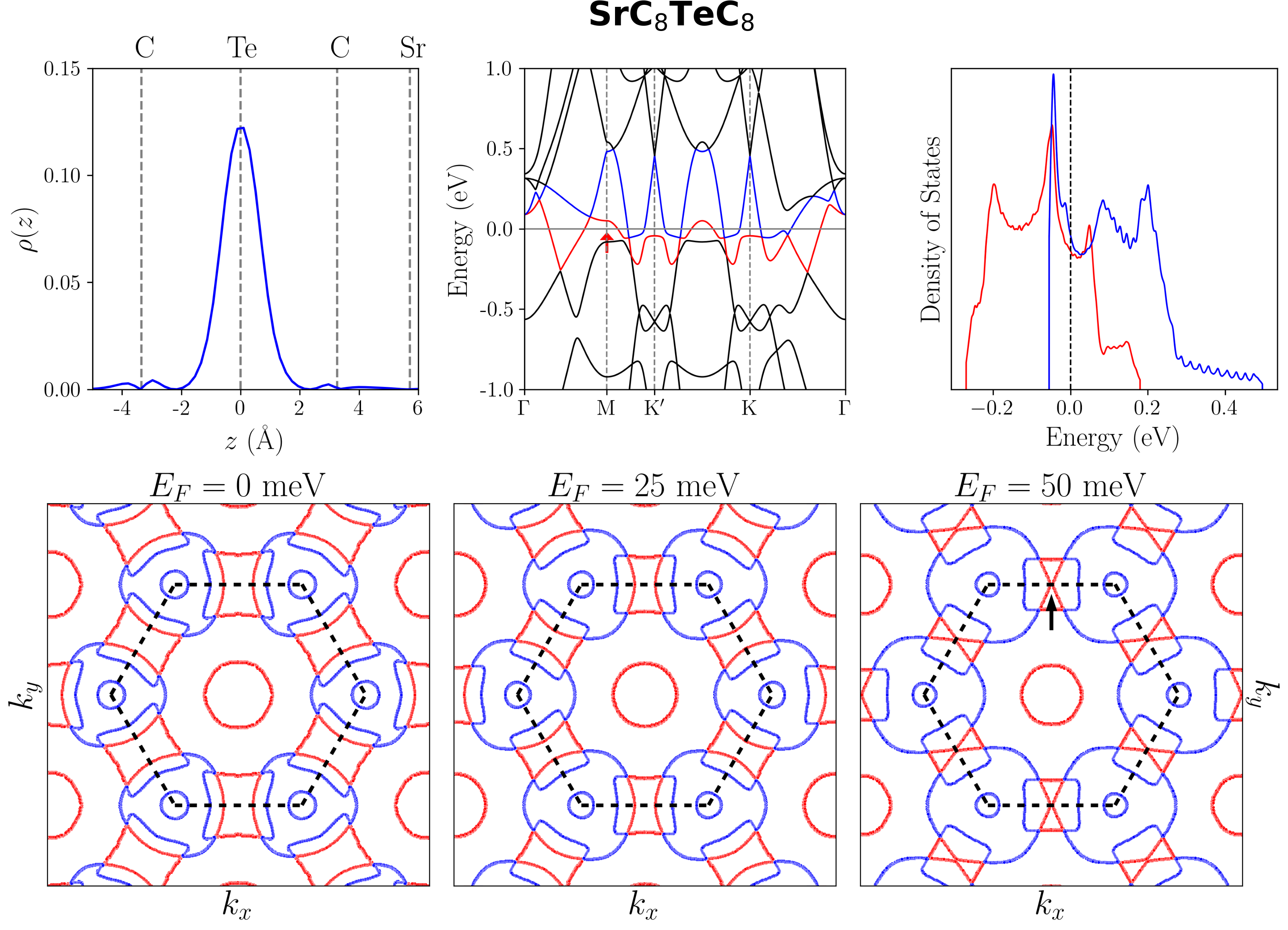}
\end{tabular}
\end{figure}

\begin{figure}[h]
\begin{tabular}{c}
\includegraphics[width=0.9\textwidth]{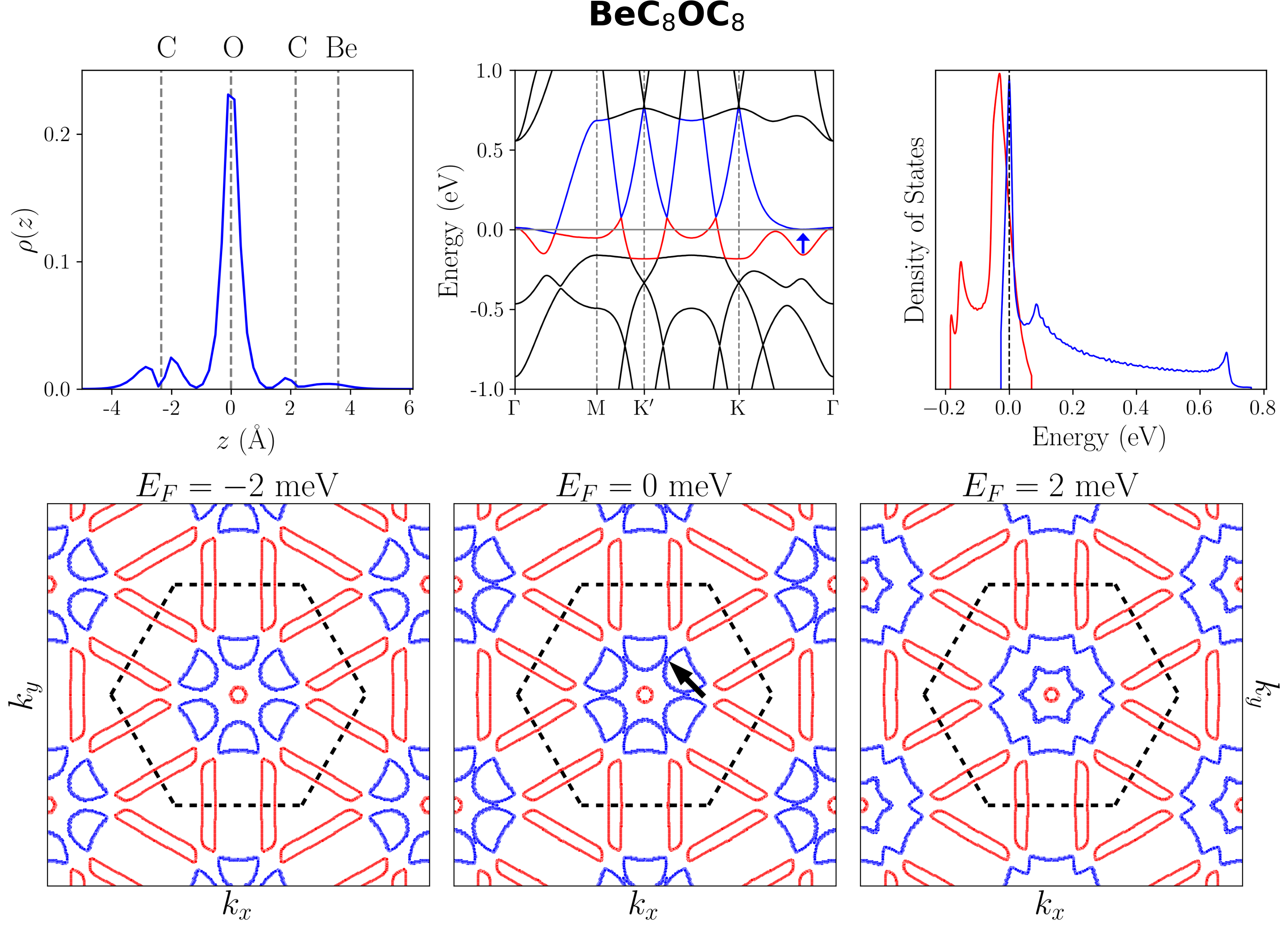}
\\
\includegraphics[width=0.9\textwidth]{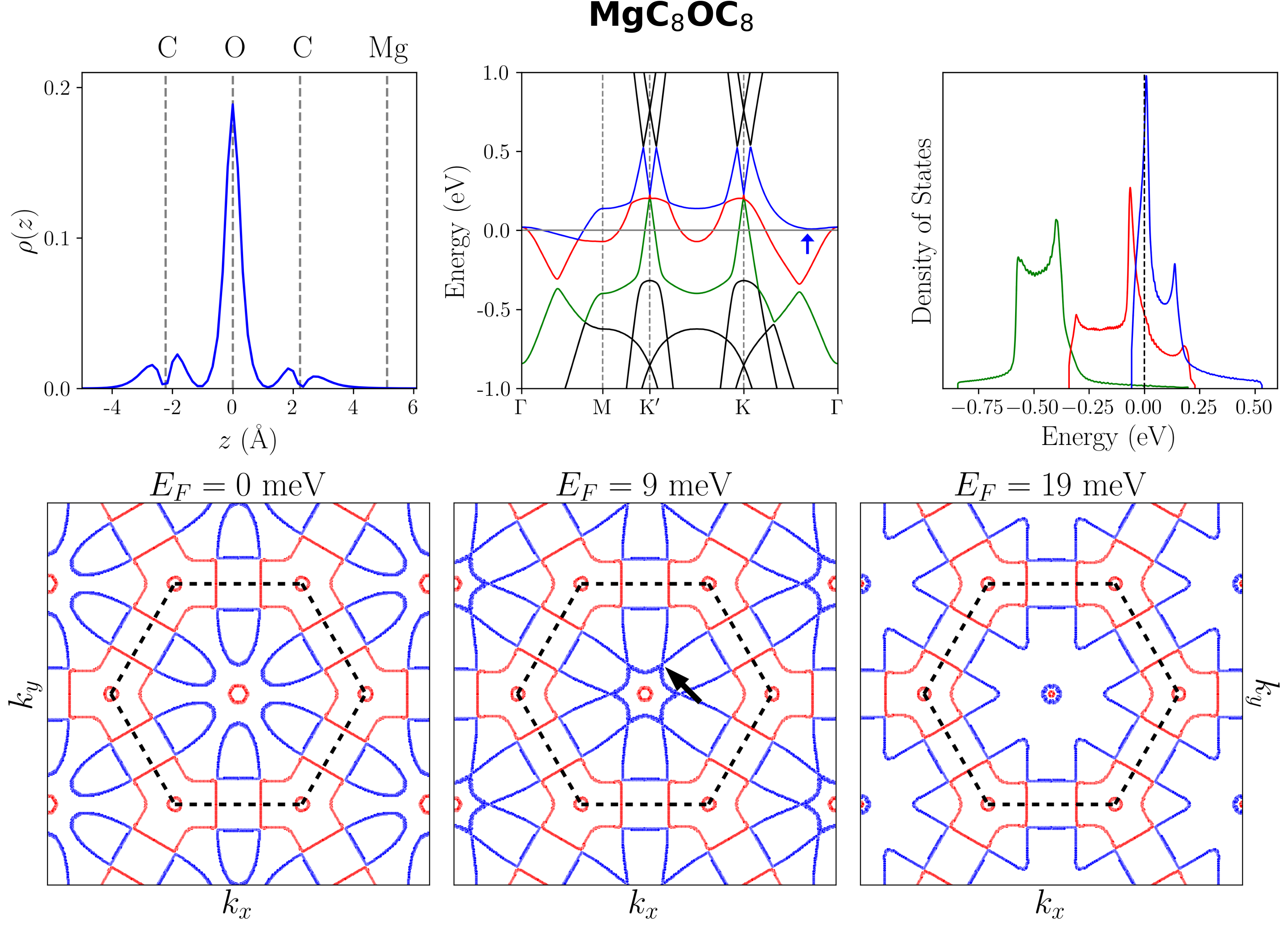}
\end{tabular}
\end{figure}

\begin{figure}[h]
\begin{tabular}{c}
\includegraphics[width=0.9\textwidth]{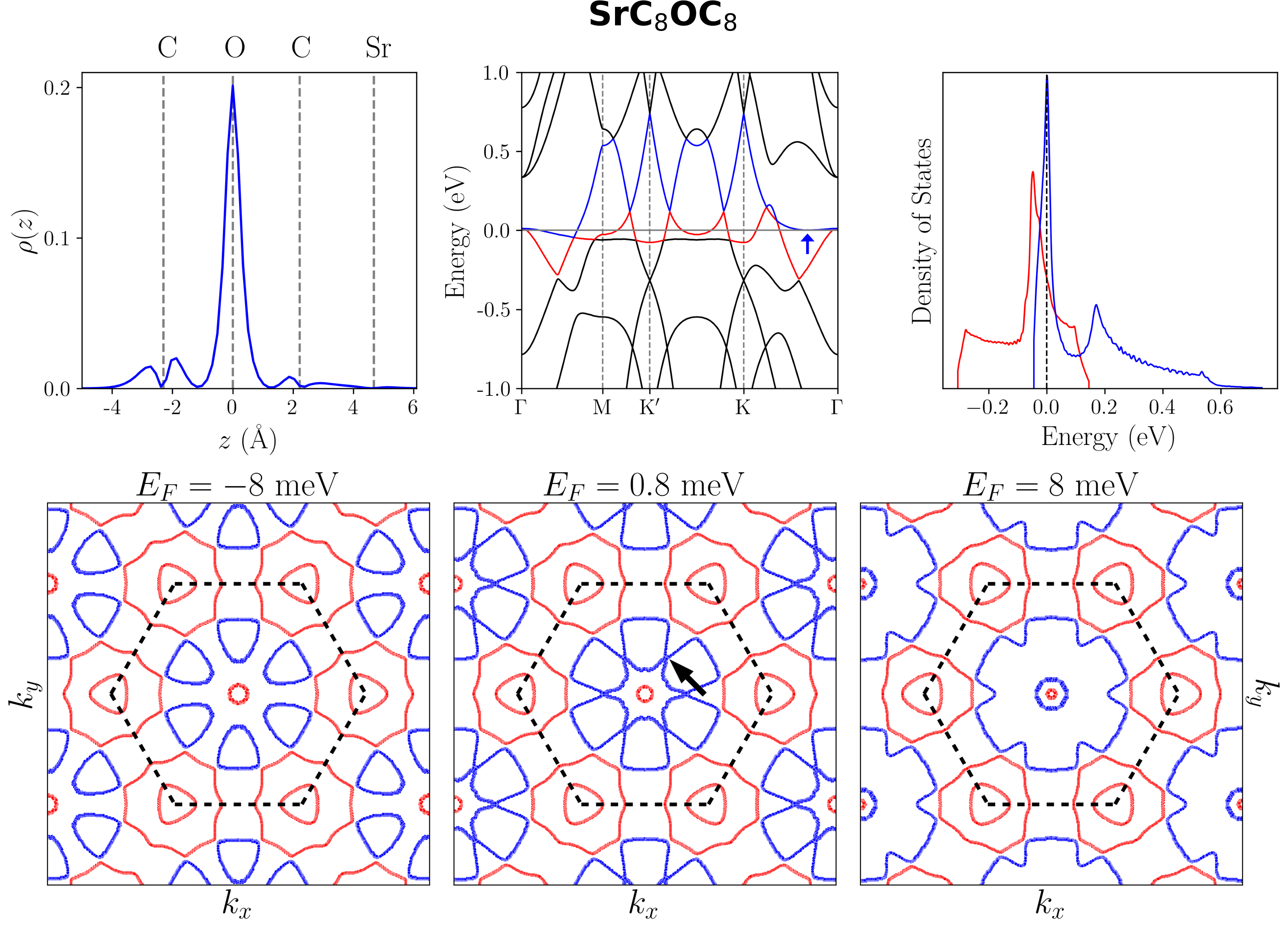}
\\
\includegraphics[width=0.9\textwidth]{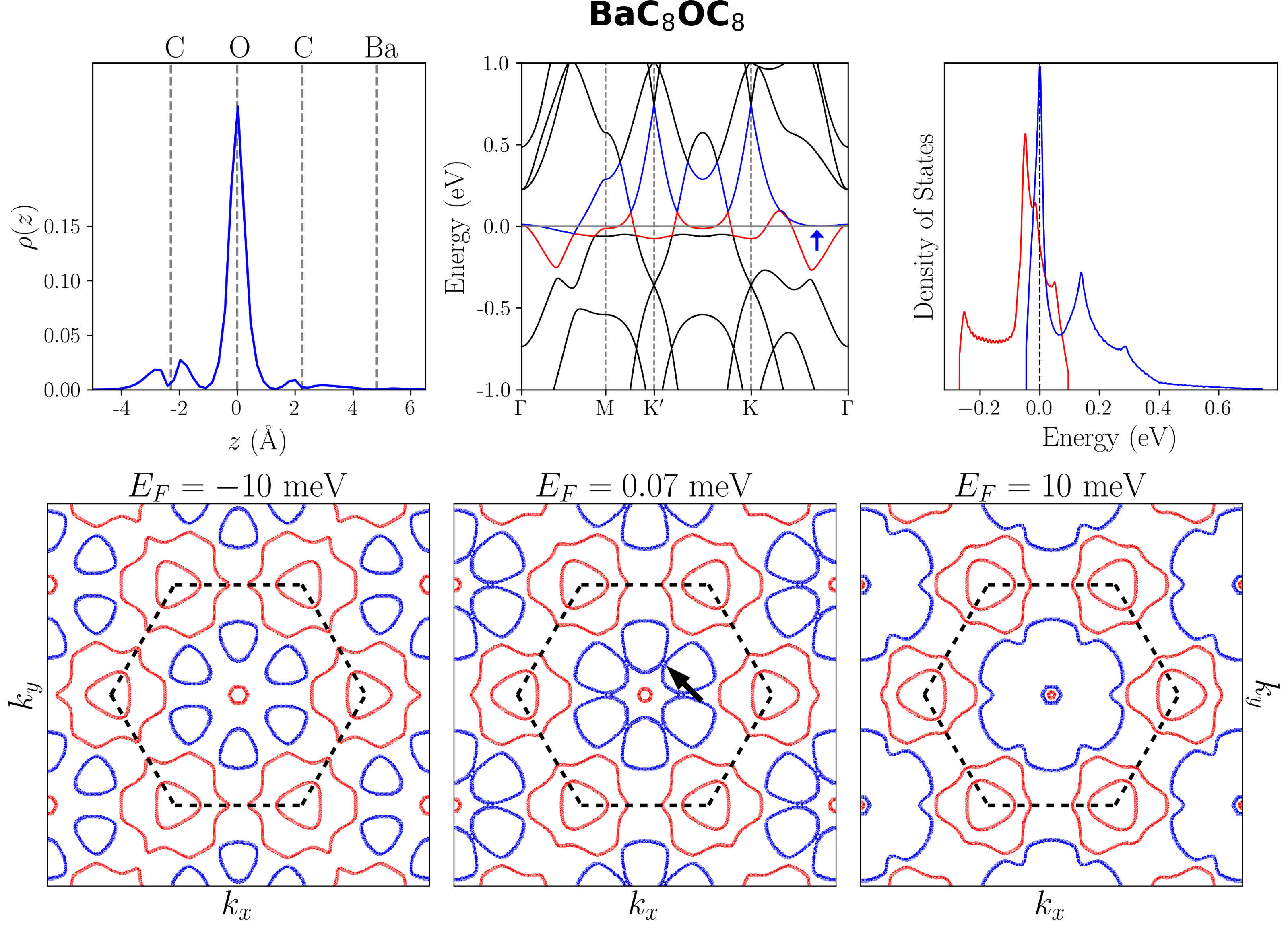}
\end{tabular}
\end{figure}

\begin{figure}[h]
\begin{tabular}{c}
\includegraphics[width=0.9\textwidth]{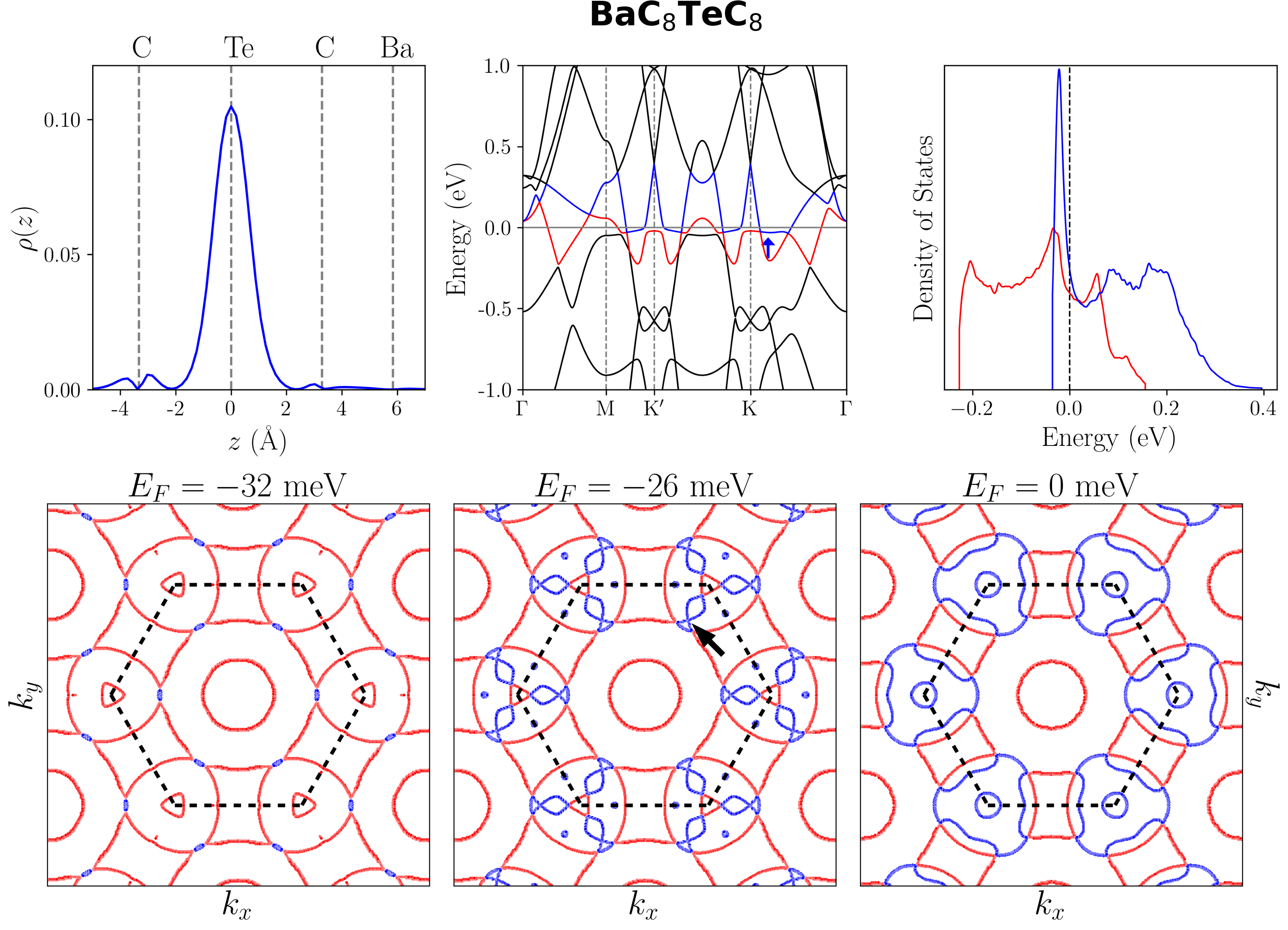}
\end{tabular}
\end{figure}